\documentclass[a4paper,11pt]{article}
\usepackage{jheppub}
\usepackage[utf8]{inputenc}
\usepackage{amsmath,amssymb,bm,slashed,braket}
\usepackage{dsfont}
\usepackage{amsfonts}
\usepackage{bm,bbm}
\usepackage{graphicx}
\usepackage{slashed} 
\usepackage[dvipsnames,table]{xcolor}
\usepackage[normalem]{ulem}
\usepackage{pifont}
\usepackage{siunitx} 
\usepackage{hyperref}
\hypersetup{colorlinks,citecolor= blue,linkcolor= blue, urlcolor=blue}
\usepackage{ulem}
\usepackage{array}
\usepackage{verbatim}
\usepackage{epsfig}
\usepackage{multirow, booktabs}
\usepackage{bbold}
\usepackage[version=4]{mhchem}
\usepackage[capitalise]{cleveref}
\usepackage{makecell}
\usepackage{orcidlink}
\usepackage{ascmac}
\usepackage{here}
\usepackage{upgreek}
\usepackage{subcaption}

\allowdisplaybreaks[4]
\newcommand{\calO}{ {\cal O} }
\newcommand{\C}{ {\tt C} }
\newcommand{\tL}{ {\tt L} }
\newcommand{\tR}{ {\tt R} }
\newcommand{\T}{ {\tt T} }

\usepackage[compat=1.1.0]{tikz-feynman}
\usepackage{subcaption}
\usetikzlibrary{positioning,arrows.meta,calc,patterns}  

\title{Light fermionic dark matter window in the scotogenic inverse seesaw model
}
\author[a,b]{Huan-Can Liang\,\orcidlink{0009-0005-6835-2496},}
\emailAdd{lhuancan@m.scnu.edu.cn}
\affiliation[a]{State Key Laboratory of Nuclear Physics and Technology,\\
Institute of Quantum Matter, South China Normal University, \\
Guangzhou 510006, China}
\affiliation[b]{Guangdong Basic Research Center of Excellence for
Structure and Fundamental \\
Interactions of Matter, Guangdong
Provincial Key Laboratory of Nuclear Science,\\
Guangzhou
510006, China}
\author[a,b]{Yi Liao\,\orcidlink{0000-0002-1009-5483},}
\emailAdd{liaoy@m.scnu.edu.cn}
\author[a,b]{Xiao-Dong Ma\,\orcidlink{0000-0001-7207-7793},}
\emailAdd{maxid@scnu.edu.cn}
\author[a,b]{Mu-Yuan Song\,\orcidlink{0000-0003-0446-5043}}
\emailAdd{muyuansong@m.scnu.edu.cn}
\author[a,b]{and Hao-Lin Wang\,\orcidlink{0000-0002-2803-5657}}
\emailAdd{whaolin@m.scnu.edu.cn}
\abstract{
The origin of neutrino mass and the nature of dark matter (DM) remain unresolved puzzles in particle physics, and an appealing possibility is to address both in a unified picture. 
This paper explores a light fermionic DM candidate within the scotogenic inverse seesaw model, which can simultaneously provide a mechanism for neutrino mass generation. 
By incorporating constraints from neutrino oscillation data, charged lepton flavor violating processes, invisible decays of the Higgs and $Z$ bosons, DM relic density, and direct detection of DM, we uncover a light fermionic DM window in the mass range $58\,{\rm GeV} \lesssim m_{\tt DM} \lesssim 63\,{\rm GeV}$ that can satisfy all of the aforementioned constraints. 
We find that this window can be jointly tested by next-generation ton-scale DM direct detection experiments including PandaX-xT and XENONnT, Higgs invisible decays, and future lepton colliders such as ILC. 
}

\makeatletter
\gdef\@fpheader{}
\makeatother
\begin{document} 

 \maketitle
\setcounter{page}{2}
\section{Introduction}

Two of the most compelling puzzles---the origin of tiny neutrino masses and the nature of dark matter (DM)---remain unsolved in modern particle physics. 
Within the Standard Model (SM), neutrinos are massless, and there is no viable DM candidate. 
However, the observed neutrino flavor oscillations and astrophysical evidence for the existence of DM provide strong motivation to extend the SM framework~\cite{Super-Kamiokande:1998kpq,SNO:2002tuh,KamLAND:2002uet,Bertone:2016nfn,Balazs:2024uyj,Marsh:2024ury}. Among various neutrino mass models, the scotogenic seesaw models and their extensions~\cite{Tao:1996vb,Ma:2006km,Krauss:2002px,Cai:2017jrq} are especially compelling, since they could simultaneously address these two big problems in a unified picture.

The original scotogenic model proposed by Tao~\cite{Tao:1996vb} and Ma~\cite{Ma:2006km} has been extensively investigated from various phenomenological perspectives; see, for example, Refs.~\cite{Schmidt:2012yg,Toma:2013zsa,Cai:2017jrq} and references therein.
However, its extension via the scotogenic inverse seesaw framework~\cite{Fraser:2014yha} remains relatively less explored, despite being equally economical. 
The main difference between the two frameworks lies in their $\mathbb{Z}_2$-odd beyond-standard-model field content: the scotogenic model introduces a second Higgs doublet and three right-handed fermion singlets, while the scotogenic inverse seesaw incorporates three real scalar singlets, a vector-like lepton doublet, and a right-handed fermion singlet. 
Due to the $\mathbb Z_2$ symmetry, the lightest $\mathbb Z_2$-odd neutral scalar or fermion can be a viable DM candidate in the scotogenic inverse seesaw model. 
In the literature, Refs.~\cite{Restrepo:2015ura,Esch:2016jyx,Liu:2025swd} studied a scalar DM candidate and hadronic collider signatures, such as di-lepton events with missing transverse energy from charged fermion production. 
On the other hand, studies in~\cite{Konar:2020wvl, Borah:2022zim} have explored fermionic DM candidates with electroweak-scale masses, considering either equal left- and right-handed Yukawa couplings related to the vector-like lepton or neglecting one of them entirely. Moreover, leptogenesis within this model has been studied in~\cite{Konar:2020vuu}. Given the ongoing and planned DM direct detection experiments~\cite{LZ:2024zvo,PandaX:2024qfu,XENON:2025vwd,PANDA-X:2024dlo,LZ:2018qzl,XENON:2020kmp} and future collider facilities such as the International Linear Collider (ILC)~\cite{Adolphsen:2013kya}, a comprehensive investigation of this model is strongly motivated. 

In this work, we investigate a fermionic DM candidate ($\chi$) in the scotogenic inverse seesaw and examine its detection prospects through both DM direct detection experiments and future lepton colliders.
Unlike previous studies~\cite{Konar:2020wvl, Borah:2022zim,Konar:2020vuu}, we do not impose specific assumptions on Yukawa couplings; instead, we adopt a more general approach by scanning the parameters within the perturbativity-allowed range. 
We derive the neutrino mass matrix and express the associated new Yukawa couplings in terms of neutrino oscillation parameters. Using experimental neutrino oscillation data as input, we perform a scan over the remaining independent model parameters, incorporating constraints from lepton flavor violating processes,
invisible decays of the Higgs and $Z$ bosons, and DM relic density requirements. 
We then project the allowed parameter space onto the plane of DM-nucleon scattering cross sections vs DM mass, identifying a viable window where the DM mass lies in the range $58\,{\rm GeV} \lesssim m_\chi \lesssim 63\,{\rm GeV}$, consistent with the current direct detection limits. 
In our analysis, we find that the spin-independent (SI) direct detection cross section is suppressed by at least four orders of magnitude compared to the spin-dependent (SD) cross sections. 
As a result, current experimental bounds on the SD cross sections impose stronger constraints on the parameter space. 
This suppression arises from the Majorana nature of the DM, which prohibits SI DM-vector-quark-current interactions. 
Interestingly, the survived parameter space can be fully tested through the SI search of next-generation direct detection experiments. 

With the DM mass being approximately half that of the Higgs mass, it can be produced directly in collider experiments. 
We conduct a complementary exploration of the allowed DM parameter space through di-lepton plus missing energy signatures at the future ILC. 
Our analysis demonstrates that only a narrow region of DM masses ($58.7\,{\rm GeV} \lesssim m_\chi \lesssim 59.3\,{\rm GeV}$) can be probed with statistical significance above $2\,\sigma$.
By selecting appropriate polarization configurations for the positron and electron beams ($P_{e^+}=+20\,\%,P_{e^-}=-60\,\%$), certain parameter points can achieve statistical significance beyond $2.5\,\sigma$.

The remainder of this paper is organized as follows. 
In \cref{sec:model}, we introduce the scotogenic inverse seesaw and identify the DM candidate after electroweak symmetry breaking. 
The neutrino mass matrix is calculated in \cref{sec:neutrinomass}. Constraints from lepton flavor violation and the invisible decays of the Higgs and $Z$ bosons are discussed in \cref{sec:lfvlimit} and \cref{subsec:hZinv}, respectively.  
The DM relic density and DM-nucleon scattering cross sections relevant to direct detection are presented in \cref{sec:dmlimit}. 
A numerical scan of the parameter space satisfying all the above constraints is described in \cref{sec:numerical-scan}.   
Collider signatures for the viable DM window at the future ILC are explored in \cref{sec:collider}. 
Finally, we conclude in \cref{sec:conclu}.

\section{The scotogenic inverse seesaw model}
\label{sec:model}

The scotogenic inverse seesaw model was originally proposed in Ref.\,\cite{Fraser:2014yha}. 
The model extends the SM particle content by an $\rm SU(2)_{\mathrm{L}}$ doublet vector-like lepton $E=\binom{E^0}{E^-}\sim(\bm{1},\bm{2},-1/2)$, a SM singlet Majorana field $N=N_R^{\tt C}+N_R\sim(\mathbf{1},\mathbf{1},0)$, and three real singlet scalar fields $\phi_i\sim(\mathbf{1},\mathbf{1},0)$ $(i=1,2,3)$, whose bare masses are denoted by $m_E$, $m_N$, and $\tilde m_{\phi_i}$, respectively. 
The three numbers in parentheses indicate how they transform with respect to the SM gauge group $\rm SU(3)_{\rm c}\times SU(2)_{\rm L}\times U(1)_{\rm Y}$.
These new fields ($E$, $N$, and $\phi_i$) are required to be odd under an imposed $\mathbb{Z}_2$ symmetry while the SM particles are $\mathbb{Z}_2$-even. 
Due to the $\mathbb{Z}_2$ symmetry,  
the model generates neutrino masses at the 1-loop level, and simultaneously provides a DM candidate, identified as the lightest $\mathbb{Z}_2$-odd particle.
The full Lagrangian is then given by
\begin{align}
\mathcal{L} 
= \mathcal{L}_{\tt SM}  +\mathcal{L}_{\tt kin.}^{\tt NP} + \mathcal{L}_{\tt Yuk.}^{\tt NP}  - V_{\tt pot.}^{\tt NP},
\end{align}
where $\mathcal{L}_{\tt SM}$ denotes the SM Lagrangian, 
and the terms involving new fields are 
\begin{subequations}
\label{eq:LagNP}
 \begin{align}
\mathcal{L}_{\tt kin.}^{\tt NP} 
&= \overline{E} (i\slashed{D}-m_E) E 
+\frac{1}{2}\overline{N} (i\slashed{\partial}-m_N) N 
+\frac{1}{2}(\partial_{\mu}\phi_p \partial^{\mu}\phi_p - \tilde m_{\phi_p}^2 \phi_p^2),
\\
-\mathcal{L}_{\tt Yuk.}^{\tt NP} 
&= y_{\phi,pr}\overline{E} L_p  \phi_r 
+ y_1 \tilde H^\dagger\overline{N} P_\tL E 
+ y_2 \tilde H^\dagger\overline{N} P_\tR E + \mathrm{h.c.},    
\\
V_{\tt pot.}^{\tt NP}  
&=
\frac{1}{2} \kappa_{pr}\, \phi_p \phi_r H^\dagger H
+ \frac{1}{4} \lambda_{\phi,prst}\, \phi_p \phi_r \phi_s \phi_t . 
\label{eq:LNP}
\end{align}
\end{subequations}
Here, $L_p$ represents the SM lepton doublets and $p,r,s,t=1,2,3$ denote the family indices.
$H$ is the SM Higgs doublet and $\tilde H =i\sigma_2 H^*$ with $\sigma_2$ being the second Pauli matrix.
$y_{\phi,pr}$ and $y_{1,2}$ are new Yukawa couplings. 
$\kappa_{pr}$ and $\lambda_{\phi,prst}$ denote new Higgs quartic couplings that are totally-symmetric among the family indices. In our analysis, without loss of generality, we take $\kappa_{pr}$ to be diagonal. The effect of a non-diagonal $\kappa_{pr}$ would lead to a redefinition of $y_{\phi,pr}$ and $\lambda_{\phi,prst}$.

After electroweak symmetry breaking (EWSB), the Higgs doublet acquires a vacuum expectation value (VEV) $H\to v/\sqrt{2}$,
and the masses of new scalars become $m_{\phi_i}^2=\tilde m_{\phi_i}^2 + \kappa_{ii}v^2/2$. 
Turning to the fermion sector, the resulting mass terms for the new neutral fermions are 
\begin{align}
-\mathcal{L}_{\rm mass}^{\rm new} 
=\frac{1}{2} \, \overline{F_\tL} \, M_{F} \, F_\tL^\C +\mathrm{ h.c.} ,
\end{align}
where
\begin{align}
F_{\tL} \equiv 
\begin{pmatrix}
E_\tL^0  \\  
E_\tR^{0\,\C}\\  
N_\tR^\C
\end{pmatrix}, \;
M_{F}= 
\begin{pmatrix}
0        & m_E       & \mu_1  \\
m_E       &  0        & \mu_2  \\
\mu_1   & \mu_2   & m_N
\end{pmatrix},
\label{MF matrix}
\end{align}
with $\mu_1 \equiv y_1 v/\sqrt{2}$ and $\mu_2 \equiv y_2 v/\sqrt{2}$. 
For simplicity, here we have assumed that the Yukawa couplings $y_1$ and $y_2$ are real.
The mass matrix $M_F$ can be diagonalized by a unitary $3\times 3$ transformation $U_F$, $U_{F} M_{F}U_{F}^{T}=M_\chi^{\rm diag}\equiv{\rm diag}(m_{\chi_1},\,m_{\chi_2},m_{\chi_3})$,
which leads to three physical Majorana states $\chi_\tL'\equiv(\chi_{1\tL},\,\chi_{2\tL},\,\chi_{3\tL})^\T = U_F F_\tL$.
Therefore, the mass term becomes 
\begin{align}
-\mathcal{L}_{\rm mass}^{\rm new} 
= \frac{1}{2} \, \overline{\chi'_\tL} \, M_{\chi}^{\rm diag} \chi_\tL'^{\C}  +\mathrm{ h.c.} 
=\frac{1}{2} m_{\chi_i} \overline{\chi_i}\chi_i
, \quad 
\chi_i \equiv \chi_{i\tL} + \chi_{i\tL}^\C.
\end{align}

The three physical masses can be identified, up to a sign difference, as the three eigenvalues of $M_F$, which can be obtained by solving its characteristic equation:
\begin{align}
x^3 -m_N x^2 -(m_E^2 +\mu_1^2 +\mu_2^2)x + m_N m_E^2 -2 \mu_1 \mu_2  m_E =0 .
\end{align}
Generally, this cubic equation has three real roots that can be written as 
\begin{subequations}
\label{eq:chimass}
\begin{align}
X_1 &= \frac{m_N}{3} +\frac{2}{3}\sqrt{m_N^2 +3(m_E^2 +\mu_1^2 +\mu_2^2)}\cos\frac{\theta}{3}  , 
\\
X_2 & =\frac{m_N}{3} +\frac{2}{3}\sqrt{m_N^2 +3(m_E^2 +\mu_1^2 +\mu_2^2)}\cos\frac{2\pi+\theta}{3},
\\
X_3 & =\frac{m_N}{3} +\frac{2}{3}\sqrt{m_N^2 +3(m_E^2 +\mu_1^2 +\mu_2^2)}\cos\frac{2\pi-\theta }{3}, 
\label{eq:dm_mass}
\end{align}
\end{subequations}
where the angle $\theta$ is defined as
\begin{align}
\theta \equiv 
\arccos\left[\frac{2m_N^3 -9m_N(2m_E^2 -\mu_1^2 -\mu_2^2) +54m_E \mu_1 \mu_2}{2[m_N^2 +3(m_E^2 +\mu_1^2 +\mu_2^2)]^{3/2}}\right] , \,
0 \leq \theta \leq \pi.
\end{align}
As $\theta$ runs from 0 to $\pi$, $X_1>0$ and $X_2<0$ always hold, but the sign of $X_3$ changes from negative to positive. 
For negative eigenvalues, the minus sign can be absorbed by a phase redefinition for the relevant Majorana fields. In our convention, we choose 
\begin{align}
m_{\chi_1} = X_1, \; m_{\chi_2} =|X_2|, \; m_{\chi_3} = |X_3|. 
\end{align}
From the above results, the unitary matrix is determined to be 
\begin{align}
U_{F}&= 
\begin{pmatrix}
1  & 0  & 0 \\
0  & e^{i\frac{\pi}{2}} & 0 \\
0  & 0 & e^{i\,H(-X_3)\frac{\pi}{2}}
\end{pmatrix}
\begin{pmatrix}
\frac{\mu_2 m_E +\mu_1 X_{1}}{\Sigma_1^2 }  & \frac{\mu_1 m_E +\mu_2 X_{1}}{\Sigma_1^2 }  & \frac{X_{1}^2 -m_E^2}{\Sigma_1^2}  \\
\frac{\mu_2 m_E +\mu_1 X_{2}}{\Sigma_2^2 }  & \frac{\mu_1 m_E +\mu_2 X_{2}}{\Sigma_2^2 }  & \frac{X_{2}^2 -m_E^2}{\Sigma_2^2}  \\
\frac{\mu_2 m_E +\mu_1 X_{3}}{\Sigma_3^2 }  & \frac{\mu_1 m_E +\mu_2 X_{3}}{\Sigma_3^2 }  & \frac{X_{3}^2 -m_E^2}{\Sigma_3^2}  
\end{pmatrix},
\end{align}
where $H(x)$ is the Heaviside step function, 
and 
\begin{align}
\Sigma_i^2 \equiv \sqrt{(\mu_2 m_E +\mu_1 X_{i})^2 +(\mu_1 m_E +\mu_2 X_{i})^2 +(X_{i}^2 -m_E^2)^2 },~
i=1,2,3.
\end{align}
As can be seen in \cref{eq:chimass}, $m_{\chi_1}\geq m_{\chi_3}$. Furthermore, $m_{\chi_2}\geq m_{\chi_3}$ when $m_E\geq m_N$. In this work, we assume that $m_E\sim m_{\phi_i} \gg m_N$, leading to $\chi_3$ as a fermionic DM candidate in our consideration. 

Regarding the charged component $E^\pm$ of the vector-like doublet, 
it does not mix with the SM charged leptons due to the exact $\mathbb{Z}_2$ symmetry. 
Consequently, $E^\pm$ remain as charged particles with a mass $m_{E^\pm} = m_E$. 
The direct search for new charged leptons from LEP collider experiments has established a lower bound on its mass of $m_E \gtrsim 100\,\mathrm{GeV}$~\cite{L3:2001xsz,ParticleDataGroup:2024cfk}.

\subsection{Neutrino masses}
\label{sec:neutrinomass}

\begin{figure}[htbp]
\centering
\includegraphics[width=1.0\linewidth]{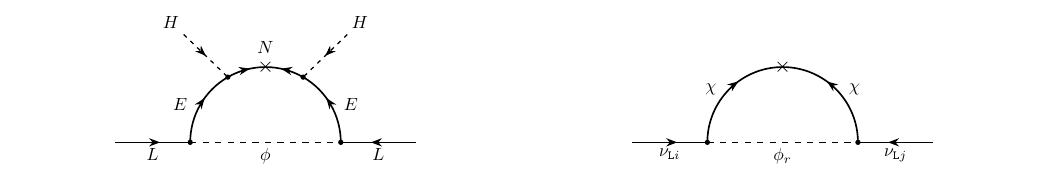}
\caption{
One-loop neutrino mass generation before  (left) and after (right) electroweak symmetry breaking.
}
\label{Fig:1loop_v_mass}
\end{figure}

Unlike the scotogenic model~\cite{Tao:1996vb,Ma:2006km}, the scotogenic inverse seesaw model generates Majorana neutrino masses exclusively through new Yukawa interactions at the one-loop level. 
The relevant Feynman diagrams before and after EWSB are shown in~\cref{Fig:1loop_v_mass}.
In the mass eigenstates, the relevant Yukawa interactions are
\begin{align}
{\cal L}_{\tt Yuk.}^{\tt NP} \supset y_{\phi,pr}\overline{E_\tR^0}\nu_{\tL,p}\phi_r
+{\rm h.c.}
= y_{\phi,pr} U_{F,k2}^* \overline{\chi_{\tL k}'^{\C}}\nu_{\tL,p}\phi_r +{\rm h.c.}.
\end{align}
Since all loop particles are much heavier than neutrinos, the neutrino mass matrix elements $M_{ij}^{\nu}$ can be obtained to excellent precision by evaluating the loop integral in \cref{Fig:1loop_v_mass} for a vanishing external momentum, yielding the result 
\begin{align}
{\cal L}^{\nu}_{\rm mass} = - \frac{1}{2}
\overline{\nu_{\tL i}^\C} M_{ij}^{\nu} \nu_{\tL j} +{\rm h.c.},
\label{eq:mass_matrix}
\end{align}
where
\begin{subequations}
\begin{align}
M_{ij}^{\nu} &=
\left( y_{\phi} \Lambda y_{\phi}^{\T} \right)_{ij},
\\
\Lambda_r &= 
\frac{1}{16\pi^2} \sum_{k=1}^{3} (U_{F,k2}^*)^2
\frac{m_{\chi_{k}}^3}{m_{\phi_r}^2 -m_{\chi_{k}}^{2}}  
\ln \frac{m_{\phi_r}^2}{m_{\chi_{k}}^2}. 
\label{eq:Lamba_r}
\end{align} 
\end{subequations}
In the above, the indices $r$ and $k$ run over the scalar and fermion mass eigenstates inside the loop, respectively.
$\Lambda = \mathrm{diag}(\Lambda_1, \Lambda_2, \Lambda_3)$ is a diagonal matrix effectively characterizing the loop functions.

The symmetric neutrino mass matrix $M^\nu$ can be diagonalized by a unitary matrix $U$, the so-called Pontecorvo–Maki–Nakagawa–Sakata (PMNS) matrix,
$U^\T M^\nu U=M^\nu_{\rm diag} = \mathrm{diag}(m_{\nu_1}, m_{\nu_2}, m_{\nu_3})$, leading to three massive Majorana neutrino states. 
Given the symmetric structure of \cref{eq:mass_matrix}, one can parameterize the Yukawa coupling matrix $y_\phi$ in terms of the physical neutrino masses and matrix $U$ by  the Casas-Ibarra parameterization~\cite{Casas:2001sr,Ibarra:2003up}
\begin{align}
y_\phi = U^* \sqrt{ M^{\nu}_{\rm diag} } R^\T \sqrt{\Lambda}^{-1},
\label{eq:y_phi}
\end{align}
where $R$ is an arbitrary complex orthogonal matrix satisfying $R R^\T = \mathbb{1}$. 
For simplicity, we set $R=\mathbb{1}$. 
To simplify the later numerical analysis, we focus only on the normal neutrino mass ordering (NO), which is currently favored by the global fitting data~\cite{Esteban:2024eli}. 
We adopt the central values of the NO oscillation parameters from the latest PDG as our inputs~\cite{ParticleDataGroup:2024cfk}: 
$\Delta m_{21}^2=7.5\times 10^{-5}\,\mathrm{eV}^2$, 
$\Delta m_{23}^2=2.45\times 10^{-3}\,\mathrm{eV}^2$, 
$\sin^2\theta_{12}=0.307$, 
$\sin^2\theta_{23}=0.534$, 
$\sin^2\theta_{13}=0.0216$, 
and $\delta_{\mathrm{CP}}=1.21\pi$.
The two Majorana CP phases in $U$ are set to zero. 
Additionally, we fix the lightest neutrino mass at $m_{\nu_1} = 0.01\,\mathrm{eV}$ as a benchmark value.
Numerically, we find that our conclusions are largely unaffected by the choices of the neutrino mass ordering, the lightest neutrino mass, and the matrix $R$.

\subsection{Lepton flavor violating processes} 
\label{sec:lfvlimit}

\begin{figure}[t]
\centering
\includegraphics[width=0.98\linewidth]{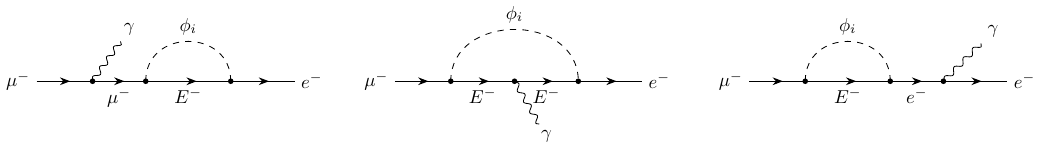}
\caption{
One-loop Feynman diagrams contributing to the CLFV process $\mu^- \to e^- \gamma$.
}
\label{mu2egamma}
\end{figure}

The new Yukawa interactions in the charged lepton sector of the model are expected to induce charged lepton flavor violating (CLFV) processes, which are subject to stringent experimental constraints. 
These processes include the two-body radiative decays $\ell_i \to \ell_j \gamma$, three-body decays $\ell_i\to \ell_j\ell_k\ell_l$, and the $\mu-e$ conversion in nuclei. 
Among these processes, the $\mu \to e \gamma$ and $\mu\to eee$ decays, and the $\mu{\rm-}e$ conversion in nuclei typically impose the most stringent constraints on the parameter space.
Currently, the MEG II and the SINDRUM experiments provide the strongest upper bounds on their occurrence, with
$\mathcal{B}(\mu^- \to e^- \gamma) < 3.1 \times 10^{-13} \quad (90\%~\text{C.L.})$~\cite{MEGII:2023ltw}, 
$\mathcal{B}(\mu^- \to e^- e^+ e^-) < 1.0 \times 10^{-12}\, (90\%~\text{C.L.})$~\cite{SINDRUM:1987nra}, 
and 
$\mathrm{CR}(\mu^- \mathrm{Au} \to e^- \mathrm{Au}) < 7 \times 10^{-13}\, (90\%~\text{C.L.})$~\cite{SINDRUMII:2006dvw}, respectively.
In the near future, the MEG II and Mu3e collaborations will reach sensitivities of $\mathcal{B}(\mu \to e \gamma) < 6.0 \times 10^{-14}$~\cite{MEGII:2023ltw} and $\mathcal{B}(\mu \to eee) < 1.0 \times 10^{-16}$~\cite{Mu3e:2020gyw}, respectively. 
For $\mu-e$ conversion in aluminum targets, the projected $90\%$ C.L. upper limit from Mu2e Run I is $\mathrm{CR}(\mu^- \mathrm{Al} \to e^- \mathrm{Al}) < 6.2 \times 10^{-16}$~\cite{Mu2e:2022ggl}, and further reaches $\mathcal{O}(10^{-17})$ by COMET Phase-II~\cite{Moritsu:2022lem}.

The relevant one-loop Feynman diagrams contributing to $\mu \to e \gamma$ are shown in \cref{mu2egamma}. Normalizing the decay width to the total width of the muon, $\Gamma_{\mu}^{\mathrm{tot}} \simeq G_{F}^{2} m_{\mu}^{5} / (192\pi^{3})$,
the branching ratio for the radiative decay becomes 
\begin{align}
  \mathcal{B} (\mu^{-} \to e^{-} \gamma )  
  &=\frac{3\alpha_{\rm em}}{64\pi G_{F}^{2}} \frac{1}{m_{E}^{4}} 
  \Big|\sum_{i} y_{\phi,1i}^* y_{\phi,2i} F\left(x_i\right) \Big |^{2} ,
  \label{eq:Brmu2egamma}
\end{align} 
where $\alpha_{\rm em}$ and $G_F$ are the fine structure constant and the Fermi constant, respectively. 
$x_i \equiv m_{\phi_{i}}^{2}/m_{E}^{2}$, and the loop function $F(x)$ is defined as
\begin{align}
    F(x) \equiv \frac{1 -6x +3x^{2} +2x^{3} -6x^{2} \ln x}{6(1-x)^{4}}.
\end{align}
It is clear from \cref{eq:Brmu2egamma} that the branching ratio is related to the neutrino oscillation parameters through the Yukawa coupling matrix $y_\phi$, as defined in \cref{eq:y_phi}. 

On the other hand, the $\mu\to eee$ decay and $\mu{\rm-}e$ conversion processes are severely suppressed in our interested parameter space. 
This can be understood as follows.
Although the relevant Feynman diagrams contributing to these processes share the same topologies as those in the scalar DM scenario discussed in Ref.~\cite{Esch:2016jyx}, 
the derived constraints differ significantly, owing to the distinct parameter space favored by DM relic density requirements.
Specifically, Ref.~\cite{Esch:2016jyx} demonstrated that in the case of scalar DM, achieving efficient annihilation often necessitates sizable Yukawa couplings $y_\phi$. 
This implies that the dominant contributions to the decay $\mu\to eee$ and $\mu{\rm-}e$ conversion are from box diagrams, with the amplitude being proportional to $y_\phi^4$, surpassing the dipole contributions ($\sim y_\phi^2 \alpha_{\rm em}$). 
In contrast, 
the fermionic DM scenario in our consideration allows for efficient annihilation via gauge interactions, requiring sizable $y_{1,2}$ to achieve the correct relic density. Consequently, small neutrino masses lead to much smaller $y_\phi$.
Numerically, we find that both $\mathcal{B}(\mu \to 3e)$ and the $\mu-e$ conversion rate are smaller by approximately two orders of magnitude relative to $\mathcal{B}(\mu \to e \gamma)$. 
This suppression aligns with the dipole-dominance relation, 
where these processes scale roughly as $\mathcal{O}(\alpha_{\mathrm{em}})$ relative to the radiative decay~\cite{Kuno:1999jp}, 
indicating that box diagram contributions are indeed subleading.

\subsection{Higgs and $Z$ boson invisible decays}
\label{subsec:hZinv}
In our considered parameter space, the lightest neutral fermion $\chi_3$ (denoted as $\chi$ hereafter) is stable and serves as the DM candidate.
For DM masses below half of the Higgs or $Z$ boson masses, these particles can decay invisibly into DM pairs. 
The interactions between DM and the Higgs and $Z$ bosons are given by:
\begin{subequations}
\label{eq:LchihZ}
\begin{align}
    \mathcal{L}_{{\rm int},\chi h} 
    &= -\frac{1}{\sqrt{2}} \left( y_1 U_{F,31} U_{F,33} +y_2 U_{F,32} U_{F,33} \right)  
     \overline{\chi} \chi h , \label{h-chi-interaction}
\\
   \mathcal{L}_{{\rm int},\chi Z}
    &=\frac{g}{4c_W} \left( \big|U_{F,31} \big|^2  -\big|U_{F,32} \big|^2 \right)
      \overline{\chi}\gamma_{\mu}\gamma_{5}\chi Z^{\mu} , \label{Z-chi-interaction}
\end{align}
\end{subequations}
where $g$ is the $SU(2)_L$ gauge coupling, and $c_W = \cos \theta_W$ with $\theta_W$ the Weinberg angle. 
The corresponding decay widths are derived from \cref{h-chi-interaction} and \cref{Z-chi-interaction} as follows
\begin{subequations}
\label{eq:hZ_inv}
\begin{align}
\Gamma \left( h\to \chi\chi\right) 
&=\frac{ (M_h^2 -4m_\chi^2)^{3/2} }{8\pi M_h^2} 
\big| y_1 U_{F,31} U_{F,33} +y_2 U_{F,32} U_{F,33} \big|^2  ,
\\
\Gamma \left( Z \to \chi\chi \right)  
&=\frac{(M_Z^2 -4m_\chi^2)^{3/2}}{12\pi}  \frac{G_F}{\sqrt{2}}
\left( \big|U_{F,31} \big|^2  -\big|U_{F,32}\big|^2 \right)^2,
\end{align}
\end{subequations}
where $M_h$ and $M_Z$ are the masses of the Higgs and $Z$ bosons.
The current experimental 90\,\% confidence level upper bounds on the branching ratios for Higgs and $Z$ invisible decays are $\mathcal{B} (h \to {\rm inv.}) < 0.107$~\cite{ATLAS:2023tkt} and $\mathcal{B} (Z \to {\rm inv.}) < 0.008$~\cite{deBoer:2021pon}.

\section{Dark matter phenomenology}
\label{sec:dmlimit} 

In this section, we investigate the phenomenology of DM, including DM production in the early Universe and detection prospects from DM direct detection experiments.

\subsection{Dark matter relic density} 
\label{subsec:dm_relic_density}

The DM relic density is determined by the thermal freeze-out mechanism through annihilation into SM particles. Its number density ($n_\chi$) is controlled by the Boltzmann equation
\begin{equation}
    \frac{dn_{\chi}}{dt} + 3Hn_{\chi} = -\langle \sigma v_{\text{M\o l}} \rangle 
    \left[ n_{\chi}^2 - (n_{\chi}^{\text{eq}})^2 \right],
    \label{eq:Boltzmann}
\end{equation}
where $H$ is the Hubble parameter and $n_{\chi}^{\text{eq}}$ is the equilibrium number density.
The quantity $\langle \sigma v_{\text{M\o l}} \rangle$ represents the thermally-averaged annihilation cross section, summed over all possible annihilation channels.
For the interested DM masses below the threshold of $W$ boson mass $m_\chi < M_W$, the annihilation channels are those into pairs of SM fermions, $\chi\chi \to f\bar{f}$, where $f$ denotes the kinematically accessible quarks and leptons. 
Among them, the $b\bar{b}$ channel typically provides the dominant contribution. 
They proceed via the s-channel diagrams mediated by the Higgs and $Z$ bosons.
$\langle \sigma v_{\text{M\o l}} \rangle$ is related to the total cross section via the relation~\cite{Gondolo:1990dk}, 
\begin{align}
    \langle \sigma v_{\text{M\o l}} \rangle \left( \chi \chi \to f \bar{f} \right)
    =\frac{4x}{K_2^2(x)} \int_{\epsilon_{\rm th}}^{\infty} 
     {\rm d} \epsilon \cdot \sigma \,\epsilon \sqrt{1+\epsilon} K_1(2x\sqrt{1+\epsilon}),
\end{align}
where $x \equiv m_\chi/T$, with $T$ being the thermal bath temperature in the early Universe, 
$K_i$ denotes the order-$i$ modified Bessel function of the second kind, $\epsilon \equiv (s -4m_\chi^2)/(4m_\chi^2)$, 
and $\epsilon_{\mathrm{th}}=\max[0, (m_f^2 - m_\chi^2)/m_\chi^2]$ is the kinematic threshold.

For the two-to-two annihilation process $\chi(p_\chi) +\chi(p_\chi')\to f(p_f) +\bar f(p_f')$, 
the spin-averaged and -summed amplitude squared takes the form
\begin{align}
 \overline{|\mathcal{M}(\chi \chi \to f \bar{f})|^2}
=& N_c^f \Big\{ 8\Big( \frac{M_Z^2}{s -M_Z^2} C_{Z,f\chi}^{\tt VA} \Big)^2 
\left[ t^2 + u^2 -2(m_{\chi}^2 -m_f^2)^2 +4m_f^2 (s- 4m_{\chi}^2) \right]  \nonumber
\\
&+8\Big( \frac{M_Z^2}{s -M_Z^2} C_{Z,f\chi}^{\tt AA} \Big)^2 
\left[ t^2 + u^2 -2(m_\chi^4 +m_f^4 -6m_{\chi}^2 m_f^2) \right]  \nonumber
\\
&+4\Big( \frac{M_h^2}{s -M_h^2} C_{h,f\chi}^{\tt SS}\Big)^2 (s- 4m_{\chi}^2 ) (s- 4m_f^2) \Big\}, 
\label{The-Amp-square}
\end{align} 
where $N_c^f = 3 (1)$ for quarks (charged leptons), and the Mandelstam variables are defined as 
\begin{align}
    s\equiv (p_\chi +p'_{\chi})^2 ,\; t\equiv (p_\chi -p_f)^2,\; 
    u\equiv (p_\chi -p_f')^2 =2(m_\chi^2 +m_f^2)-s-t .
\end{align}
The coefficients $C_{h,f\chi}^{\tt SS},\,C_{Z,f\chi}^{\tt VA,AA}$ are provided in \cref{eq:coeff_exp} in the next subsection.
Thus, the total cross section is calculated to be
\begin{align}
\sigma \left( \chi\chi \to f \bar{f} \right)  
= \frac{1}{64\pi s} \frac{1}{\bm{p}_{\chi,{\rm cm}}^2 }  \int_{t_-}^{t_+} {\rm d} t 
 \quad\overline{|\mathcal{M}(\chi \chi \to f \bar{f})|^2},
\end{align}
where $|\bm{p}_{\chi,{\rm cm}}| =(s/4 - m_\chi^2)^{1/2}$ is the DM momentum in the center-of-mass frame, and the integration limits are given by
$t_\pm =-(|\bm{p}_{\chi,{\rm cm}}| \mp |\bm{p}_{f,{\rm cm}}|)^2$ with $|\bm{p}_{f,{\rm cm}}| =(s/4 - m_f^2)^{1/2}$. 

In our numerical analysis, the Boltzmann equation in \cref{eq:Boltzmann} is solved numerically using the \texttt{MadDM} package~\cite{Arina:2020kko}, which incorporates all subleading annihilation and co-annihilation channels. 
The calculation also includes proper treatment of kinematic thresholds and resonance effects, particularly for $m_\chi \simeq M_h/2$ or $m_\chi \simeq M_Z/2$. 
We require the predicted DM relic density to agree with the Planck measurement $\Omega_{\rm Planck} h^2 = 0.120 \pm 0.001$~\cite{Planck:2018vyg} to identify the allowed parameter space.
To account for theoretical and observational uncertainties, a slightly wider range, $0.115 \leq \Omega_{\tt DM} h^2 \leq 0.125$, is adopted in our numerical analysis.
We find that the analytical calculation for $\langle \sigma v_{\text{M\o l}} \rangle$ matches the output of \texttt{MadDM}, further validating our calculation.

\subsection{Dark matter direct detection}

Now we turn to direct detection searches. 
Since DM couples to SM quarks and leptons through the exchange of the Higgs and $Z$ bosons, it can induce signals in DM direct detection experiments.
Since the typical momentum transfer in these experiments is below a few MeV, it is convenient to describe these interactions within the effective field theory (EFT) framework. 
By integrating out the Higgs and $Z$ bosons from \cref{eq:LchihZ}, the relevant leading-order effective Lagrangian reads:
\begin{align}
    \mathcal{L}_{\rm eff} 
    = \sum_f \left[ C_{h,f\chi}^{\tt SS} (\bar{f}f)(\bar{\chi}\chi)
    + C_{Z,f\chi}^{\tt VA} (\bar{f}\gamma^\mu f)(\bar{\chi}\gamma_\mu\gamma_5\chi)
    + C_{Z,f\chi}^{\tt AA} (\bar{f}\gamma^\mu\gamma_5 f)(\bar{\chi}\gamma_\mu\gamma_5\chi) \right],
    \label{eq:Leff_DMDD}
\end{align} 
where $f$ runs over the light quarks $u,d,s$ and the electron $e$. 
The matching coefficients are given by:
\begin{subequations}
\label{eq:coeff_exp}
\begin{align}
    C_{h,f\chi}^{\tt SS} 
    &= \frac{1}{\sqrt{2}v}\frac{m_f}{M_h^2}
    \left( y_1 U_{F,31} U_{F,33} +y_2 U_{F,32} U_{F,33} \right) , 
    \label{eq:coeff_SS} 
\\
    C_{Z,f\chi}^{\tt VA} 
    &= \frac{G_F}{\sqrt{2}}
    \left( |U_{F,31}|^2  - |U_{F,32}|^2 \right) g_V^f , 
    \label{eq:coeff_VA}
\\
    C_{Z,f\chi}^{\tt AA} 
    &= -\frac{G_F}{\sqrt{2}}
    \left( |U_{F,31}|^2  - |U_{F,32}|^2 \right) g_A^f ,
    \label{eq:coeff_AA}
\end{align}
\end{subequations}
where $1/v=2^{1/4}G_F^{1/2}$.
The vector and axial-vector couplings of the $Z$ boson to the SM fermions are defined as 
\begin{align}
g_V^f =T_f^3 -2Q_f s_W^2, \quad
g_A^f =T_3^f ,
\end{align}
with $(T_3^f, Q_f ) =(1/2, 2/3),\, (-1/2, -1/3 ),\,(-1/2, -1 )$ for the up- and down-type quark, and the electron, respectively.
Given the comparable coupling strengths of DM-electron and DM-quark ($u,d$) interactions, the electron recoil signal is significantly suppressed compared to that of the nucleon recoil for DM in the mass range of tens GeV due to kinematic limitations. 
As a result, it does not impose meaningful constraints within our considered mass range.
We therefore neglect DM-electron scattering effects and focus exclusively on DM-nucleon scattering in the following analysis.

It is a standard practice to match the above effective Lagrangian onto the nonrelativistic (NR) EFT framework, where the relevant ingredients are the unity and spin operators of the DM and nucleon (including the proton and neutron), 
to calculate the DM-nucleus scattering rate. 
For the effective Lagrangian in \cref{eq:Leff_DMDD}, the matched NREFT Lagrangian takes the form
\begin{align}
{\cal L}_{\rm NREFT} 
= \sum_{\textsc n=p,n} \left( c_1^\textsc n \mathcal{O}_1^\textsc n + c_4^\textsc n \mathcal{O}_4^\textsc n 
+ c_6^\textsc n \mathcal{O}_6^\textsc n  + c_8^\textsc n \mathcal{O}_8^\textsc n + c_9^\textsc n \mathcal{O}_9^\textsc n \right) , 
\label{eq:NREFT}
\end{align}
where the relevant NREFT operators and matching coefficients are defined as
\begin{subequations}
\begin{align}
\mathcal{O}_1^\textsc n =& \mathbb{1}, &
c_1^\textsc n =& 8 m_\chi m_{\textsc n} \sum_q C_{h,q\chi}^{\tt SS} F_{S}^{q/\textsc n}(0) ,
\\
\mathcal{O}_4^\textsc n =& \bm{s}_\chi \cdot \bm{s}_\textsc n, &
c_4^\textsc n  =& -32 m_\chi m_\textsc n \sum_q C_{Z,q\chi}^{\tt AA} G_A^{q/ \textsc n}(0) ,
\\
\mathcal{O}_6^\textsc n =& \left( \bm{s}_\chi \cdot \bm{q} \right) \left( \bm{s}_\textsc n \cdot \bm{q} \right), &
c_6^\textsc n  = & -32 m_\chi m_\textsc n \sum_q C_{Z,q\chi}^{\tt AA} 
\Big( \frac{a_{\pi}^{q/\textsc n} }{q^2 -m_\pi^2}  +\frac{a_{\eta}^{q/\textsc n} }{q^2 -m_\eta^2} \Big) , 
\\
\mathcal{O}_8^\textsc n =& \bm{s}_\chi \cdot \bm{v}_\textsc n^{\perp}, &
c_8^\textsc n  = & 16 m_\chi m_\textsc n \sum_q C_{Z,q\chi}^{\tt VA} F_1^{q/\textsc n}(0) , 
\\
\mathcal{O}_9^\textsc n =& i \bm{s}_\chi \cdot \left( \bm{s}_\textsc n \times \bm{q} \right), &
c_9^\textsc n  =& -16 m_\chi \sum_q C_{Z,q\chi}^{\tt VA} \big( F_1^{q/\textsc n}(0) + F_2^{q/\textsc n}(0) \big).
\end{align}
\end{subequations}
Here, $m_{\textsc n,\pi,\eta}$ denote the masses of the nucleon $\textsc n=p,\,n$, and the $\pi^0$ and $\eta$ mesons. 
$\bm{v}_{\textsc n}^{\perp}$ is the DM-nucleon transverse velocity in the NR limit, and $\bm{q} \equiv \bm{k}_\chi -\bm{k}_\chi'$ represents the three-momentum transfer, with $\bm{k}_\chi (\bm{k}_\chi')$ denoting the momentum of incoming (outgoing) DM in the scattering. The DM and nucleon spin operators are denoted by $\bm{s}_\chi$ and $\bm{s}_\textsc n$, respectively. The label $q$ sums over the $u,d,s$ quark flavors. $F_{S,1,2}^{q/\textsc n},~G_A^{q/\textsc n},~a_{\pi,\eta}^{q/\textsc n}$ are the nucleon form factors, whose numerical values adopted in our analysis are summarized in \cref{tab:nucleon_FF}.
The operators $\mathcal{O}_{1,8}^\textsc n$ are related to spin-independent (SI) interactions, while $\mathcal{O}_{4,6,9}^\textsc n$ generate spin-dependent (SD) contributions.

\begin{table}[t]
\centering
\resizebox{0.8\linewidth}{!}{
\renewcommand{\arraystretch}{1}
\begin{tabular}{|c|ccc|ccc|}
\hline
\multirow{2}{*}{$F_S^{q/\textsc n}=\frac{m_\textsc n}{m_q}f_{Tq}^{(\textsc n)}$~\cite{Ellis:2018dmb}} & $f_{Tu}^{(p)}$ & $f_{Td}^{(p)}$ & $f_{Ts}^{(p)}$ & $f_{Tu}^{(n)}$ & $f_{Td}^{(n)}$ & $f_{Ts}^{(n)}$ \\
\cline{2-7}
  & $0.018$ & $0.027$ & $0.037$ & $0.013$ & $0.040$ & $0.037$ \\
\hline
\multirow{2}{*}{$F_{1}^{q/\textsc n}(0)$~\cite{Bishara:2017pfq}} &\multicolumn{2}{c}{$F_{1}^{u/p}=F_{1}^{d/n}$}  &\multicolumn{2}{c}{$F_{1}^{d/p}=F_{1}^{u/n}$}  &\multicolumn{2}{c|}{$F_{1}^{s/\textsc n}$} \\
\cline{2-7}
   &\multicolumn{2}{c}{$2$}   &\multicolumn{2}{c}{$1$}   &\multicolumn{2}{c|}{$0$}  \\
\hline
\multirow{2}{*}{$F_{2}^{q/\textsc n}(0)$~\cite{Bishara:2017pfq}} &\multicolumn{2}{c}{$F_{2}^{u/p}=F_{2}^{d/n}$}  &\multicolumn{2}{c}{$F_{2}^{d/p}=F_{2}^{u/n}$}  &\multicolumn{2}{c|}{$F_{2}^{s/\textsc n}(0)$} \\
\cline{2-7}
   &\multicolumn{2}{c}{$1.609$}   &\multicolumn{2}{c}{$-2.097$}   &\multicolumn{2}{c|}{$-0.064$}  \\
\hline
\multirow{2}{*}{$G_A^{q/\textsc n}(0)$~\cite{Lin:2018obj}}  &\multicolumn{2}{c}{$G_A^{u/p}=G_A^{d/n}$} &\multicolumn{2}{c}{$G_A^{d/p}=G_A^{u/n}$}  &\multicolumn{2}{c|}{$G_A^{s/\textsc n}$}  \\
\cline{2-7}
  &\multicolumn{2}{c}{$0.777$}  &\multicolumn{2}{c}{$-0.438$} &\multicolumn{2}{c|}{$-0.053$}  \\
\hline
\multirow{2}{*}{$a_{\pi}^{q/\textsc n}$~\cite{DelNobile:2021wmp}}  &\multicolumn{2}{c}{$a_{\pi}^{u/p} = a_{\pi}^{d/n}$}   &\multicolumn{2}{c}{$a_{\pi}^{d/p} = a_{\pi}^{u/n}$}  &\multicolumn{2}{c|}{$a_{\pi}^{s/\textsc n}$} \\
\cline{2-7}
    &\multicolumn{2}{c}{$ (G_A^{u/p} + G_A^{d/p})/2 $}   &\multicolumn{2}{c}{$-a_{\pi}^{u/p}$}  &\multicolumn{2}{c|}{$0$}  \\
\hline
\multirow{2}{*}{$a_{\eta}^{q/\textsc n}$~\cite{DelNobile:2021wmp}}  &\multicolumn{2}{c}{$a_{\eta}^{u/\textsc n}$}   &\multicolumn{2}{c}{$a_{\eta}^{d/\textsc n}$}  &\multicolumn{2}{c|}{$a_{\eta}^{s/\textsc n}$} \\
\cline{2-7}
    &\multicolumn{2}{c}{$(G_A^{u/p} +G_A^{d/p} -2G_A^{s/p})/6$}   &\multicolumn{2}{c}{$a_{\eta}^{u/\textsc n}$}  &\multicolumn{2}{c|}{$-2 a_{\eta}^{u/\textsc n}$}  \\
\hline
\end{tabular}}
\caption{Nucleon form factors employed in the numerical analysis. }
\label{tab:nucleon_FF}
\end{table}

The NREFT results allow us to determine the equivalent zero-momentum-transfer SI ($\overline{\sigma}^{\rm SI}$) and SD ($\sigma^{\rm SD}_\textsc n$) cross sections, which direct detection experiments typically adopt when setting their limits.
For the SI interactions, $\overline{\sigma}^{\rm SI}$ is defined as the coherent sum over nucleons within the nucleus, weighted by the natural isotopic abundance~\cite{Feng:2011vu}: 
\begin{align}
\overline{\sigma}^{\rm SI} 
=& \frac{4 \mu_{\chi\textsc n}^2}{\pi} 
\frac{\sum_{i}\xi_{i} \left\{ \sum_q C_{h,q\chi}^{\tt SS} \big[ Z  F_S^{q/p}
+ (A_{i}-Z) F_S^{q/n}\big] \right\}^2 \mu_{\chi T_i}^2}
{\sum_{i}\xi_{i} A_{i}^2 \mu_{\chi T_i}^2} \nonumber
\\
&+\frac{2 \mu_{\chi\textsc n}^2}{\pi}  
\frac{\sum_{i}\xi_i \left\{ \sum_{q} C_{Z,q\chi}^{\tt VA}\big[ Z  F_1^{q/p} 
+ (A_i -Z) F_1^{q/n} \big]\right\}^2 \mu_{\chi T_i}^2}
{\sum_{i}\xi_i A_{i}^2 \mu_{\chi T_i}^2} \langle v_{\rm rel}^2 \rangle , 
\label{SI-sigma}
\end{align}
where $Z$ and $A_i$ are the atomic and mass numbers of the target nucleus isotope $T_i$, whose abundance and mass are denoted by $\xi_i$ and $m_{T_i}$, respectively. For xenon-based experiments such as XENON and PandaX,  we adopt the natural abundances provided in~\cite{DelNobile:2021wmp}. 
The reduced masses between the DM and the nucleon or the nucleus $T_i$ are defined as
\begin{align}
    \mu_{\chi\textsc n} =\frac{m_\chi m_{\textsc n}}{m_\chi +m_{\textsc n}} ,\quad 
    \mu_{\chi T_i} =\frac{m_\chi m_{T_i}}{m_\chi + m_{T_i}}. 
\end{align} 
For the velocity-dependent term, $\langle v_{\rm rel}^2 \rangle =\int v^2 f(v) {\rm d} v \simeq (370 \,{\rm km/s})^2$, where the numerical estimate was obtained by using the Maxwell-Boltzmann velocity distribution for $f(v)$~\cite{Lewin:1995rx}. Our later numerical analysis indicates that the velocity-dependent term is suppressed by approximately two orders of magnitude relative to the velocity-independent contribution mediated by the Higgs boson. 

From \cref{eq:NREFT}, the SD DM-nucleon cross section due to $\calO_{4,6,9}^{\textsc n}$ is calculated to take the form 
\begin{align}
\sigma^{\rm SD}_\textsc n 
&=\frac{\mu_{\chi\textsc n}^2}{256\pi m_\chi^2 m_{\textsc n}^2} \bigg\{ 3(c_4^{\textsc n})^2 
+4\mu_{\chi\textsc n}^2 (c_9^{\textsc n})^2  \langle v_{\rm rel}^2 \rangle
\nonumber
\\
&\quad 
+ \frac{1}{ |\hat t_0| }
\int_{\hat{t}_0}^{0}  {\rm d}q^2 \big[
(c_6^\textsc n)^2 (q^2)^2 - 2 c_4^{\textsc n}
c_6^{\textsc n} q^2\big]
\bigg\},
\end{align}
where $\hat t_0 \equiv -4\bm{k}_{\chi,{\rm cm}}^2 \approx -4 \mu_{\chi\textsc n}^2 \langle v_{\rm rel}^2 \rangle$ with $\bm{k}_{\chi,{\rm cm}} \simeq \mu_{\chi\textsc n} \bm{v}_{\rm rel}$ being the DM momentum in the center-of-mass frame. 
Numerically, the first term in the curly bracket dominates over the others by at least four orders of magnitude due to velocity suppression of the latter. 
After taking into account the matching coefficients and retaining only the dominant contribution, we obtain
\begin{align} 
\sigma^{\rm SD}_\textsc n 
\approx& \frac{12\mu_{\chi\textsc n}^2}{\pi} \big| C_{Z,q\chi}^{\tt AA} G_A^{q/ \textsc n} \big|^2
\nonumber
\\
=& \frac{3G_F^2\mu_{\chi\textsc n}^2}{2\pi} 
( |U_{F,31}|^2  - |U_{F,32}|^2)^2
(G_A^{u/ \textsc n}-G_A^{d/ \textsc n}
-G_A^{s/ \textsc n})^2, 
\label{eq:SD-sigma}
\end{align}
where summation over $q$ is implied, and in the second step, we have used \cref{eq:coeff_AA}
and the values $g_A^{u,d,s}=1/2,~-1/2,~-1/2$. 
As $G_A^{s/ \textsc n} \ll |G_A^{u(d)/ \textsc n}|$ as is shown in \cref{tab:nucleon_FF}, this leads to the approximate isospin-symmetric relation $\sigma^{\rm SD}_p\approx \sigma^{\rm SD}_n$.

\section{Numerical analysis} 
\label{sec:numerical-scan}

In this section, we perform a comprehensive numerical scan of the model parameter space to identify allowed regions compatible with 
neutrino oscillation data in \cref{sec:neutrinomass} and the DM relic density in \cref{subsec:dm_relic_density}, and at the same time, satisfying the current 
experimental constraints from the $\mu\to e\gamma$ in \cref{sec:lfvlimit} and the Higgs and $Z$ boson invisible decays in \cref{subsec:hZinv}. 
Our input new physics parameters are $\{m_E, m_N, y_{1,2}, m_{\phi_{1,2,3}}\}$. 
Note that the DM mass $m_\chi$ is uniquely determined by the above parameters through \cref{eq:dm_mass}.
First, since the masses of new scalars $\phi_{1,2,3}$ only enter into the neutrino mass and LFV observables through loop functions, they have a mild influence on other observables. 
Thus, we fix their masses to $m_{\phi_1} = 1000\,\mathrm{GeV}$, $m_{\phi_2} = 1200\,\mathrm{GeV}$, and $m_{\phi_3} = 1400\,\mathrm{GeV}$, to reduce the number of scanned parameters. 
Second, for the remaining four parameters, we randomly generate approximately $1.5\times 10^6$ points, constraining their values to fall within the following ranges, 
\begin{align}
300\,\mathrm{GeV}\leq m_E \leq 1500\,\mathrm{GeV},\quad
1\,\mathrm{GeV} \leq m_N \leq 300\,\mathrm{GeV},\quad -0.5 \leq y_{1,2}\leq 0.5.
\end{align} 
We choose a lower limit of $m_E^{\rm min} =300\,\mathrm{GeV}$  to avoid exclusion limits from recent LHC searches for heavy charged particles~\cite{ParticleDataGroup:2024cfk,L3:2001xsz}. The values of the couplings $y_{1,2}$ also satisfy perturbativity constraints. 

For each set of generated parameters, we first calculate the corresponding Yukawa couplings $y_{\phi,pr}$ using \cref{eq:y_phi} based on neutrino oscillation data. We then retain the parameter set only if it satisfies all of the following conditions simultaneously:
\begin{subequations}
\label{eq:select_con}
\begin{align}
\mathcal{B}(\mu\to e\gamma) &<  3.1 \times 10^{-13},
\\
\mathcal{B}(h \to {\rm inv.}) &<  0.107,
\\
\mathcal{B}(Z \to {\rm inv.}) &<  0.008,
\\
0.115 \leq  \Omega_{\tt DM} h^2 &\leq  0.125.
\end{align}
\end{subequations}
The expressions and detailed calculations for these constraints are provided in \cref{sec:lfvlimit,subsec:hZinv,subsec:dm_relic_density}, 
respectively.
After applying the selection criteria in \cref{eq:select_con}, approximately $10^4$ data points remain from the original $1.5\times 10^6$. 
We find that the branching ratio of $\mu\to e\gamma$ is smaller by approximately six orders of magnitude compared to the current experimental limit for the remaining points. 
Based on these surviving points, we calculate the DM mass via \cref{eq:dm_mass} and determine the DM-nucleon cross sections using \cref{SI-sigma,eq:SD-sigma}.

\begin{figure}[t]
\centering     
\includegraphics[width=0.48\linewidth]{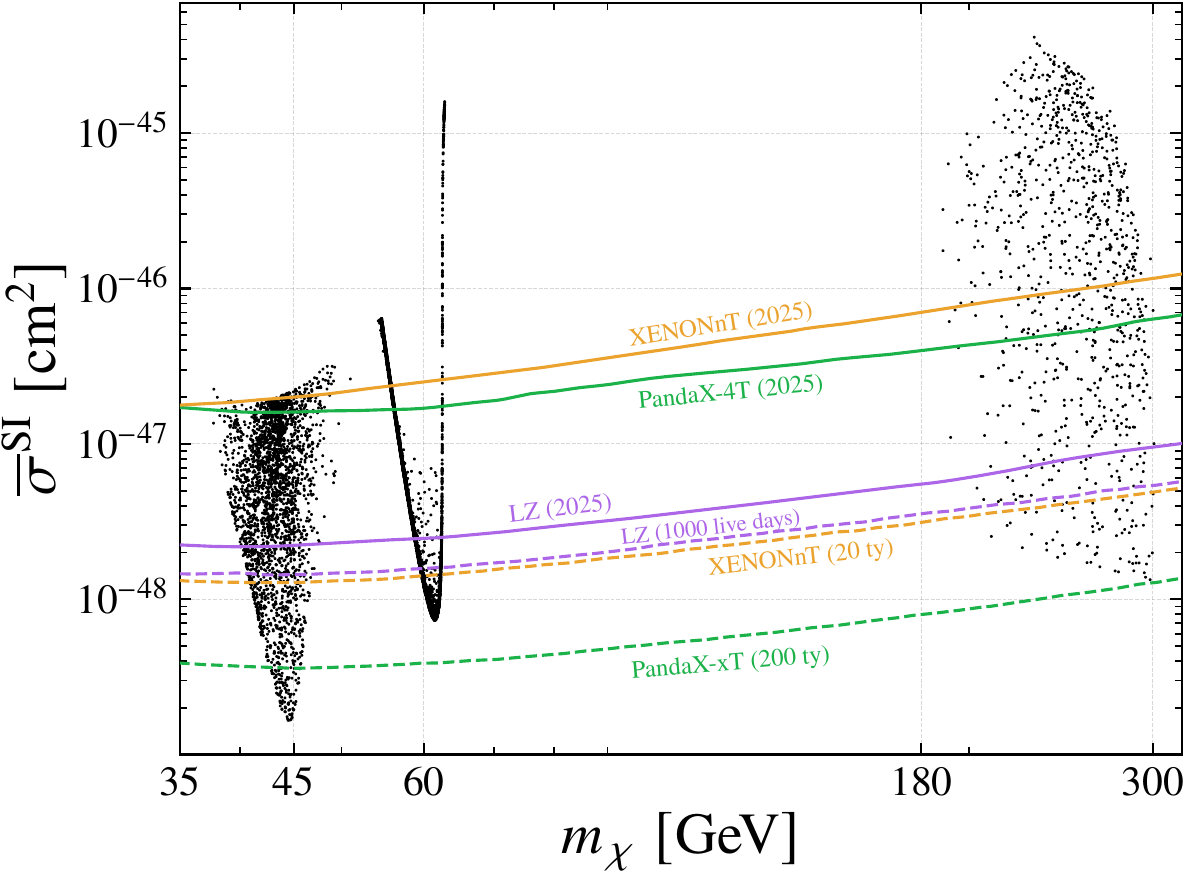}
\includegraphics[width=0.48\textwidth]{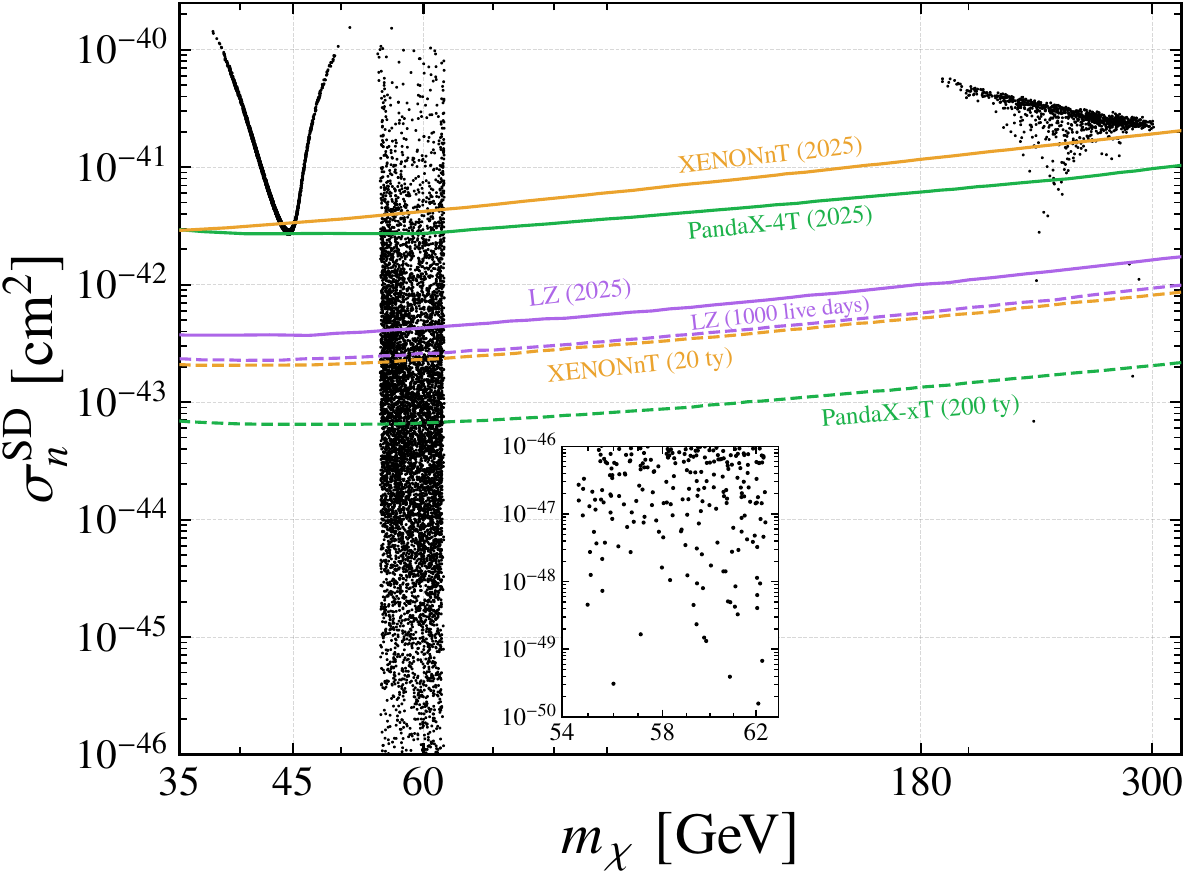}
\caption{ 
The left (right) panel shows the allowed parameter points projected onto the plane of SI (SD) DM-nucleon scattering cross section versus DM mass.
The solid orange, blue, and purple curves show current $90\%$ C.L. exclusion limits
from XENONnT~\cite{XENON:2025vwd}, PandaX-4T~\cite{PandaX:2024qfu},
and LZ~\cite{LZ:2024zvo}, respectively.
The corresponding dashed curves represent their future sensitivity projections from XENONnT~\cite{XENON:2020kmp}, 
PandaX-xT \cite{PANDA-X:2024dlo}, 
and LZ~\cite{LZ:2018qzl}. }
\label{fig:sigma_chiN}
\end{figure}

In \cref{fig:sigma_chiN}, we present the SI (left panel) and SD (right panel) DM-nucleon cross sections as functions of DM mass for the surviving points that satisfy the conditions in \cref{eq:select_con}. 
The solid and dashed curves indicate the current DM direct detection limits and future projected sensitivities of the XENON~\cite{XENON:2025vwd,XENON:2020kmp}, PandaX~\cite{PandaX:2024qfu,PANDA-X:2024dlo}, and LZ~\cite{LZ:2024zvo,LZ:2018qzl} experiments, respectively. 
We do not show the SD DM-proton scattering case because of the approximate isospin-symmetric relation $\sigma^{\rm SD}_n \approx \sigma^{\rm SD}_p$ and the weaker experimental bounds on $\sigma^{\rm SD}_p$.
Since DM is primarily composed of the neutral lepton $N$, its mass is approximately equal to $m_N$.
As shown in the figure, the viable parameter points are concentrated mainly in three regions: 
the $Z$-boson resonance region near $m_\chi \sim M_Z/2$, 
the Higgs resonance region around $m_\chi \sim M_h/2$ , 
and the high-mass region with $180\,\mathrm{GeV} \lesssim m_\chi \lesssim 300\,\mathrm{GeV} $.
This pattern is dictated by the relic density requirement. In the resonance regions, the thermally averaged annihilation cross section is enhanced due to $s$-channel propagators when $m_\chi$ approaches half the mediator mass. 
A characteristic ``dip'' structure appears near $m_\chi \sim M_{Z,h}/2$ in both the SI and SD cross sections. 
This arises because, to prevent DM over-annihilation, the effective coupling $C_{Z,f\chi}^{\tt VA,AA}\,(C_{h,f\chi}^{\tt SS})$ must decrease as $m_\chi$ approaches $M_{Z(h)}/2$, leading to suppression in the direct detection cross sections.
The upper bounds on the cross sections in these two resonance regions are constrained by the invisible decays of the Higgs and $Z$ bosons.
For the high-mass region, it is uniquely determined by the DM relic density via additional DM annihilation channels, such as $\chi\chi \to W^+W^-$ and $ZZ$.

\begin{figure}[t]
\centering
\includegraphics[width=0.48\textwidth]{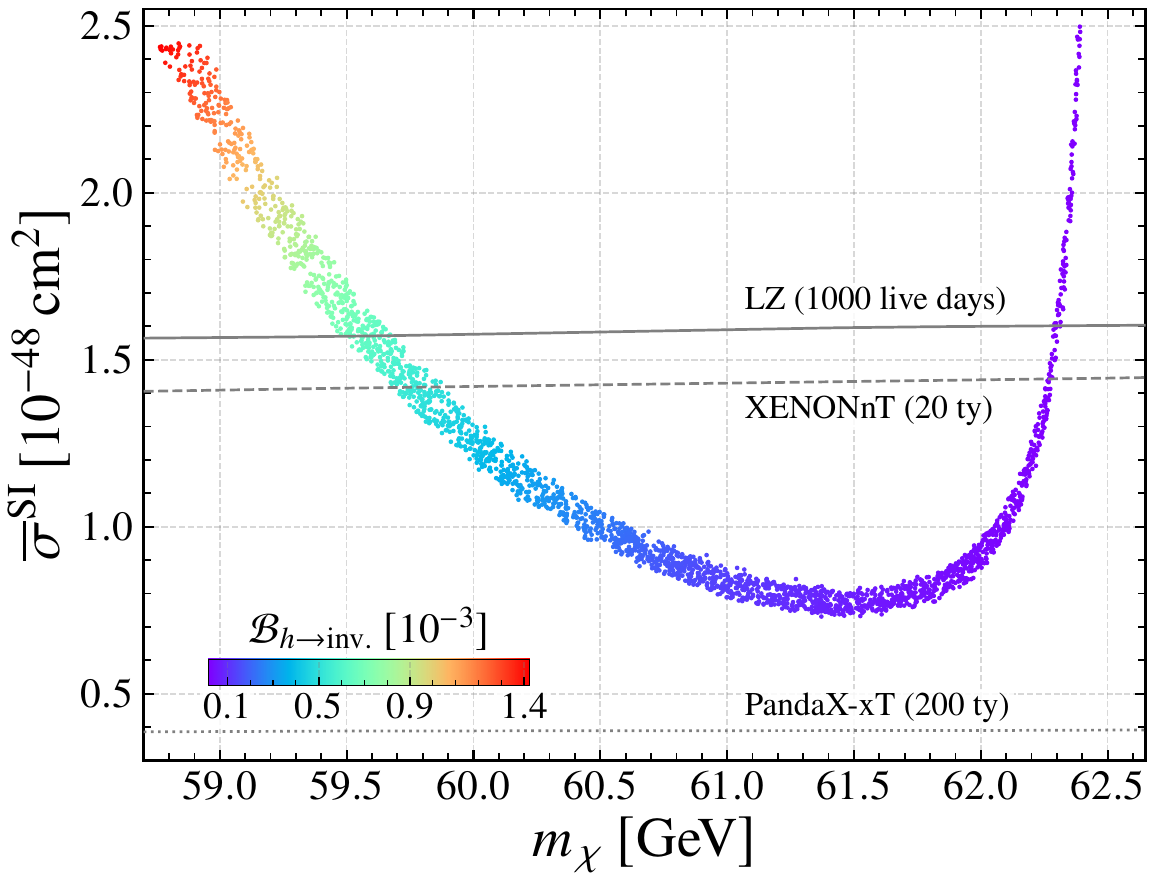}
\includegraphics[width=0.478\textwidth]{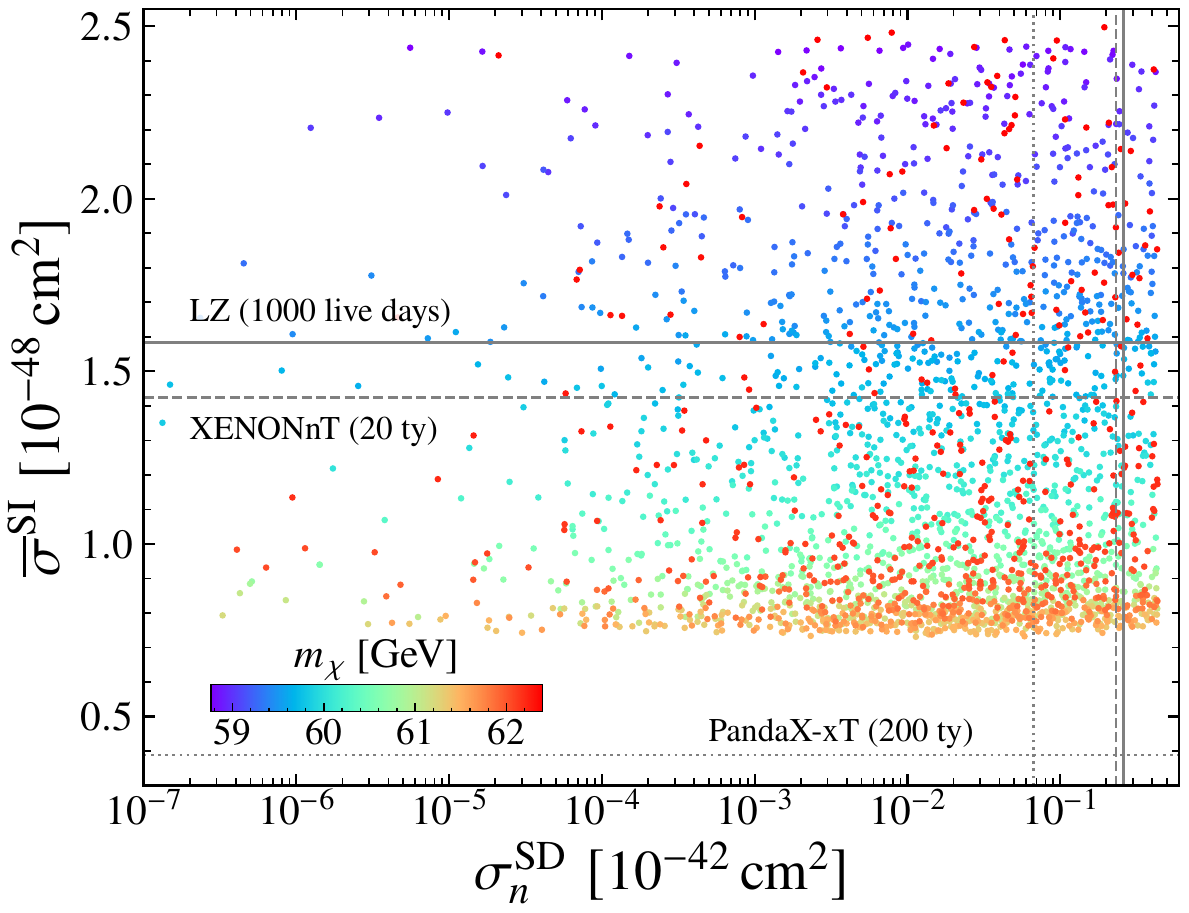}
\caption{
Distributions of allowed parameter points satisfying all current experimental constraints in the $m_\chi$-$\overline{\sigma}^{\rm SI}$ (left panel) 
and $\sigma^{\rm SD}_n$-$\overline{\sigma}^{\rm SI}$ (right panel)  planes. 
The horizontal and vertical gray lines represent the future direct detection sensitivities for the SI and SD cross sections, respectively.}
\label{fig:SISDnALL}
\end{figure}

Note that in both the $Z$ resonance region ($m_\chi\sim M_Z/2$) and the high-mass region ($m_\chi>180\,\rm GeV$), the SD cross section is typically about five orders of magnitude larger than the SI cross section. 
This enhancement occurs because SD interactions are dominantly mediated by the $Z$ boson exchange, whereas the $Z$ boson's contribution to the SI cross section is suppressed by the DM velocity effect due to its Majorana nature, as shown in \cref{SI-sigma}. 
In addition, the SI cross section mediated by the Higgs boson is further suppressed due to its dependence on the light quark masses.
From \cref{fig:sigma_chiN}, it is clear that the current LZ limits on the SD DM-neutron cross section $\sigma_n^{\rm SD}$ nearly exclude the entire $Z$ resonance region and the high-mass region. 
By combining the current experimental constraints on both SI and SD cross sections, we find that the Higgs resonance region ($58\,\mathrm{GeV} \lesssim m_\chi \lesssim 63\,\mathrm{GeV}$) remains the only viable window consistent with all relevant experimental constraints. 
\cref{fig:SISDnALL} presents the distributions of this final window in the $m_\chi$-$\overline\sigma^{\rm SI}$ (left panel) and $\sigma^{\rm SD}_n$-$\overline\sigma^{\rm SI}$ (right panel) planes, together with the projected sensitivities of future direct detection experiments: 
LZ (1000 live days)~\cite{LZ:2018qzl} (gray solid curve), XENONnT (20 ty)~\cite{XENON:2020kmp} (gray dashed curve), 
and PandaX-xT (200 ty)~\cite{PANDA-X:2024dlo} (gray dotted curve). 
It is clear that the future PandaX-xT experiment can probe the entire allowed parameter space via SI interactions, rendering this window fully testable. 
As shown in the left panel, it can also be probed through the Higgs invisible decay, with a branching ratio reaching the ${\cal O}(10^{-3})$ level~\cite{Abidi:2025mdw}.

\begin{figure}[t]
\centering
\includegraphics[width=0.48\textwidth]{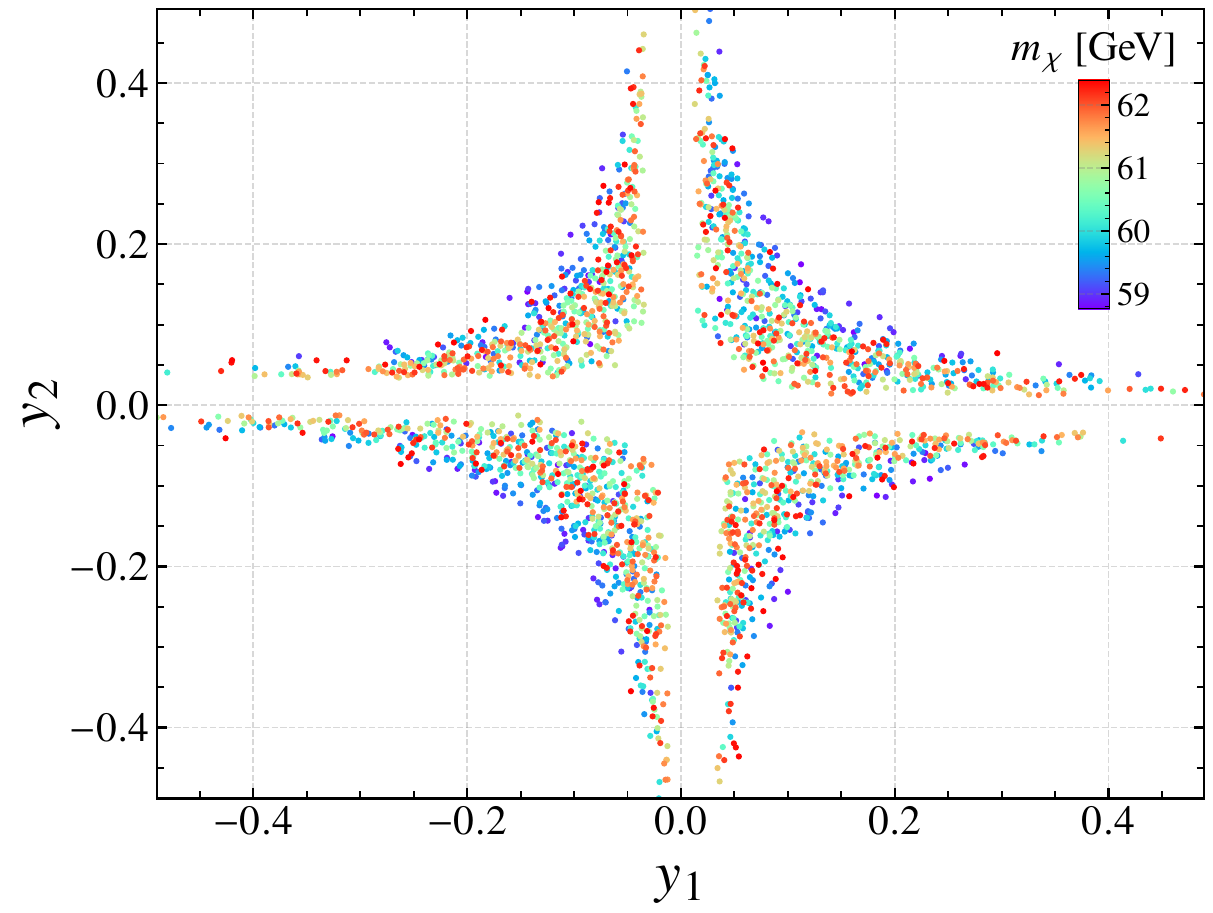}
\includegraphics[width=0.48\textwidth]{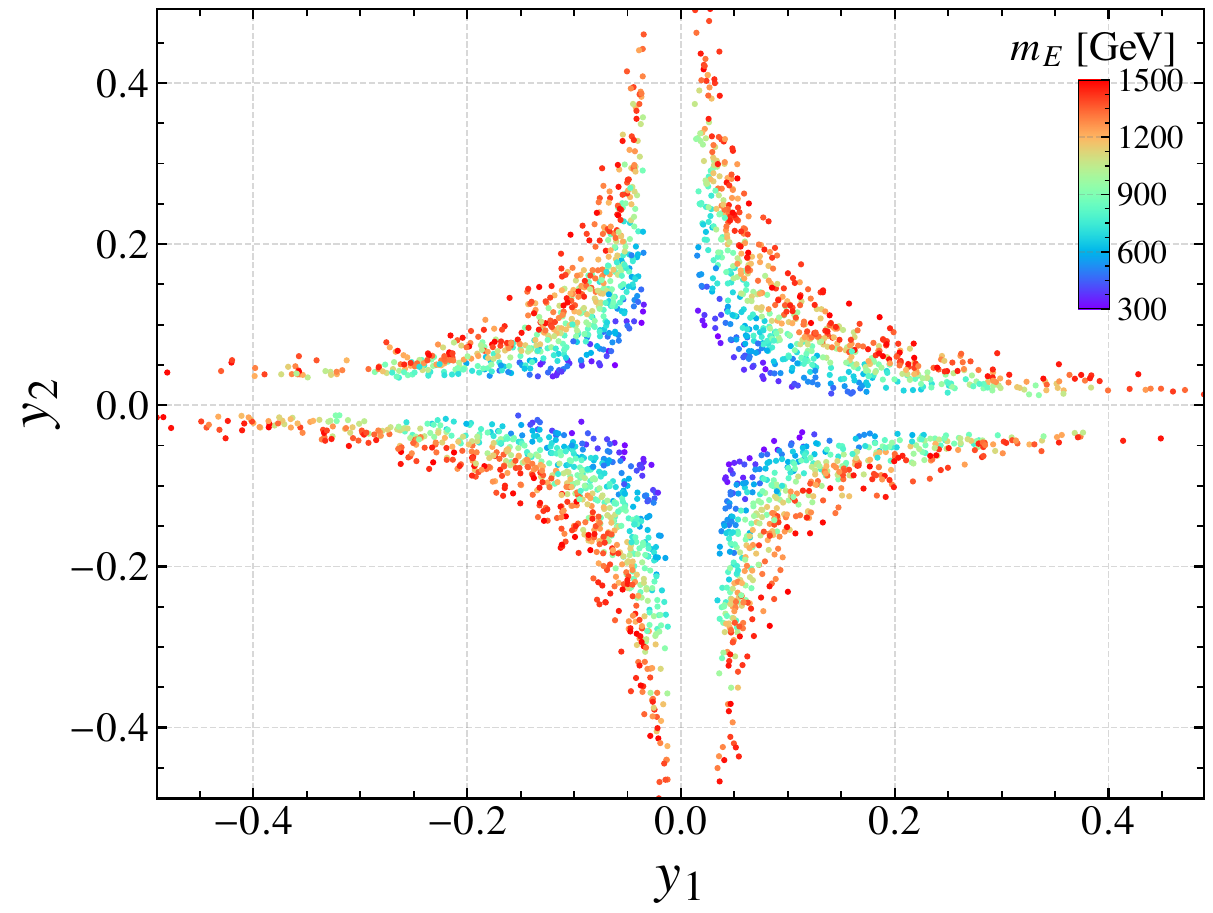}
\caption{
Allowed parameter points in the $y_1$-$y_2$ plane, satisfying all current experimental constraints. The color scale in the left (right) panel represents the corresponding DM mass $m_\chi$ (heavy vector-like lepton mass $m_E$).}
\label{fig:y1_y2_plane}
\end{figure}

Finally, \cref{fig:y1_y2_plane} shows the allowed points in the $y_1$-$y_2$ plane of the Yukawa couplings, where the color scales denote the masses of the DM and the heavy vector-like lepton in the left and right panels, respectively. 
The white regions, corresponding to either small or large $|y_{1}y_2|$, lead to DM overabundance or underabundance and are therefore excluded by DM relic density constraints.
Note that the allowed Yukawa couplings $y_1$ and $y_2$ are around $\mathcal{O}(10^{-1})$, 
in sharp contrast to the scalar DM case considered in Ref.~\cite{Esch:2016jyx}, where they remain below $\mathcal{O}(10^{-4})$ to suppress the loop-induced neutrino masses. 
In our scenario, these couplings drive DM annihilation via singlet-doublet mixing and must be sizable enough to yield the correct DM relic density. 
In the following section, we will examine the discovery potential of the allowed DM window at future lepton colliders. 

\section{Dark matter at lepton collider}
\label{sec:collider}

Collider searches for DM typically focus on missing energy signatures accompanied by visible SM particles. 
For example, the LHC targets such signals through mono-photon, mono-gluon, mono-jet, or di-lepton events produced along with invisible DM particles~\cite{Abercrombie:2015wmb, Penning:2017tmb,Boveia:2018yeb,Carpenter:2012rg,Bell:2012rg}. 
On the other hand, future high-energy lepton colliders, such as ILC~\cite{Adolphsen:2013kya},  CLIC~\cite{Aicheler:2018arh}, CEPC~\cite{CEPCStudyGroup:2023quu}, and FCC-ee~\cite{FCC:2018evy}, can extend this strategy beyond the LHC sensitivity due to lower background levels. 
The cleaner environment and well-defined initial states at lepton colliders provide a promising avenue for probing DM candidates~\cite{Birkedal:2004xn,Fox:2011fx,Wan:2014rhl,Ge:2023wye,Antusch:2025lpm}.

For the viable DM parameter space in the scotogenic inverse seesaw model that we identified in the previous section, we investigate its collider signatures at the future ILC, taking advantage of its higher center-of-mass energy $\sqrt{s}=1$ TeV.
We implement the model into \texttt{FeynRules~2.0}~\cite{Alloul:2013bka} to generate the Universal FeynRules Output (\texttt{UFO}) formats~\cite{Degrande:2011ua}, 
which is imported into \texttt{MadGraph5$\_$aMC@NLO}~\cite{Alwall:2014hca} to simulate the signal and background events.
The generated events are then analyzed using \texttt{MadAnalysis5} for detailed kinematic region selection and signal-background studies~\cite{Conte:2012fm}. 
In our analysis, we employ two distinct search channels: the mono-photon final state and the di-lepton final state with missing energy signatures, corresponding to the processes $e^+ e^- \to {E}_{\text{miss}}(\equiv \chi\chi) + \gamma$ and $e^+ e^- \to\ell^+\ell^-+{E}_{\text{miss}}$, respectively.

\begin{figure}[t]
\centering
\includegraphics[scale=0.55]{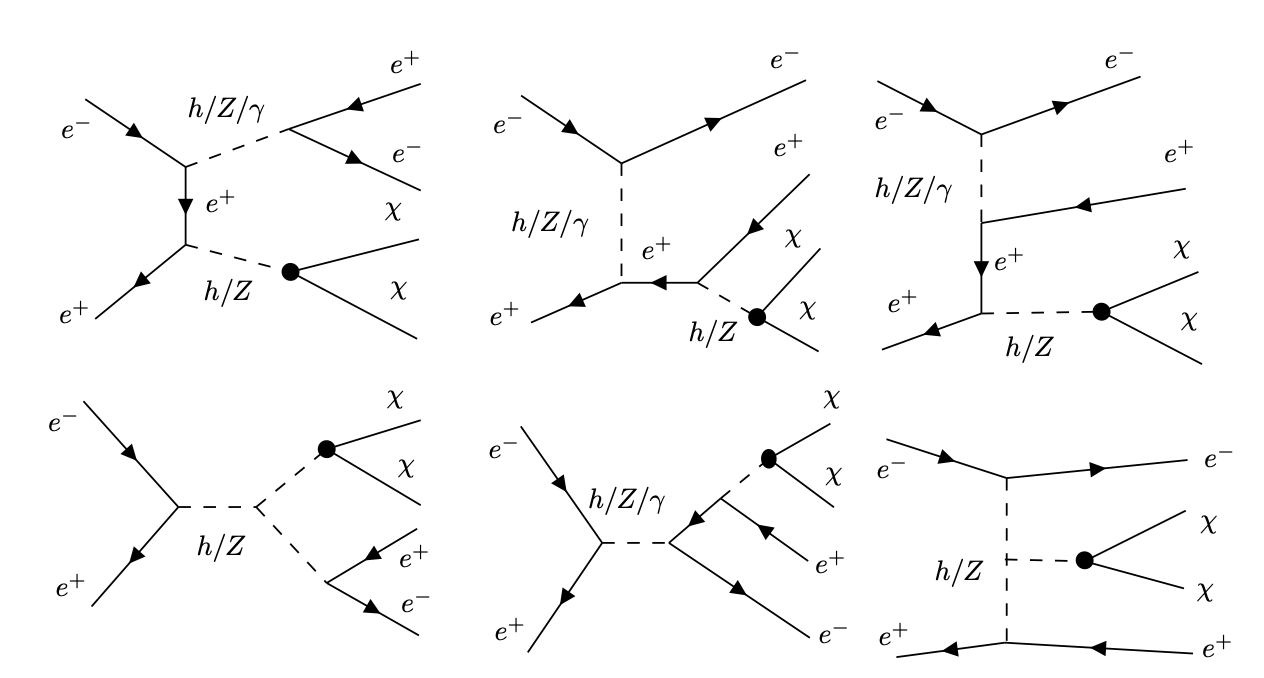}  
\caption{Representative Feynman diagrams involving a single new physics vertex (black dots) for di-electron plus DM pair production at an $e^+e^-$ collider. The dashed lines represent possible Higgs, $Z$, or $\gamma$ propagators, and the interaction points involving three dashed lines correspond to the $ZZh$ and $hhh$ vertices. 
The same diagrams apply to di-muon channels by replacing the final-state $e^+e^-$by $\mu^+\mu^-$. }
\label{fig:dileptonsignal}
\end{figure}

\begin{figure}[t]
\centering
\includegraphics[scale=0.55]{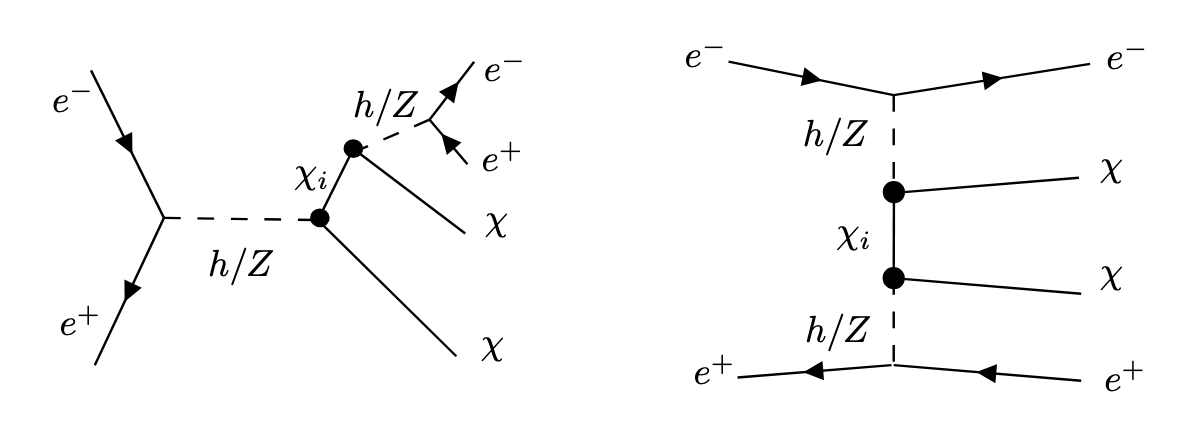} 
\caption{ The two additional Feynman diagrams for di-lepton processes involving two new physics vertices (black dots). Here, $\chi_i$ collectively denotes all three new fermion components $\chi_{1,2}$ and the DM $\chi$. }
\label{fig:dileptonsignal2}
\end{figure}

The mono-photon process $e^+e^- \to \chi\chi+\gamma$ is mainly produced via the $s$-channel exchange of the SM Higgs or $Z$ boson, with the photon emitted from initial-state radiation. The dominant background arises from $e^+e^- \to \bar{\nu}\nu + \gamma$ and $e^+e^- \to e^+e^- + \gamma$. In the latter case, the final-state $e^+e^-$ travel along the beam axis and escape detection. 
In contrast, the di-lepton process $e^+ e^- \to \ell^+ \ell^-  \chi\chi$ produces two visible leptons accompanied by missing energy signatures. The representative Feynman diagrams are shown in~\cref{fig:dileptonsignal} and \cref{fig:dileptonsignal2}, involving one and two new physics vertices, respectively.
The dominant backgrounds, as previously identified in Ref.~\cite{Yu:2014ula}, are categorized into three processes: $e^+ e^- \to \ell^+ \ell^- \bar{\nu}\nu$ (with $\ell=e,~\mu$), $e^+ e^- \to \tau^+\tau^-\bar{\nu}\nu$ and $e^+ e^- \to \tau^+\tau^-$.
The processes $e^+ e^- \to \ell^+ \ell^- \bar{\nu}\nu$ and $e^+ e^- \to \tau^+\tau^-\bar{\nu}\nu$ are mediated by either $t$-channel lepton exchange or $s$-channel Higgs decays to $W^+W^-/ZZ$ boson pairs, with subsequent decays to the di-lepton final states. The process $e^+ e^- \to \tau^+\tau^-$ involves direct production of a tau pair via $s-$channel processes. The tau leptons in these backgrounds are identified through their leptonic decays into muons or electrons, accompanied by neutrinos.

\begin{figure}[t!]
\centering
\includegraphics[width=0.48\textwidth]{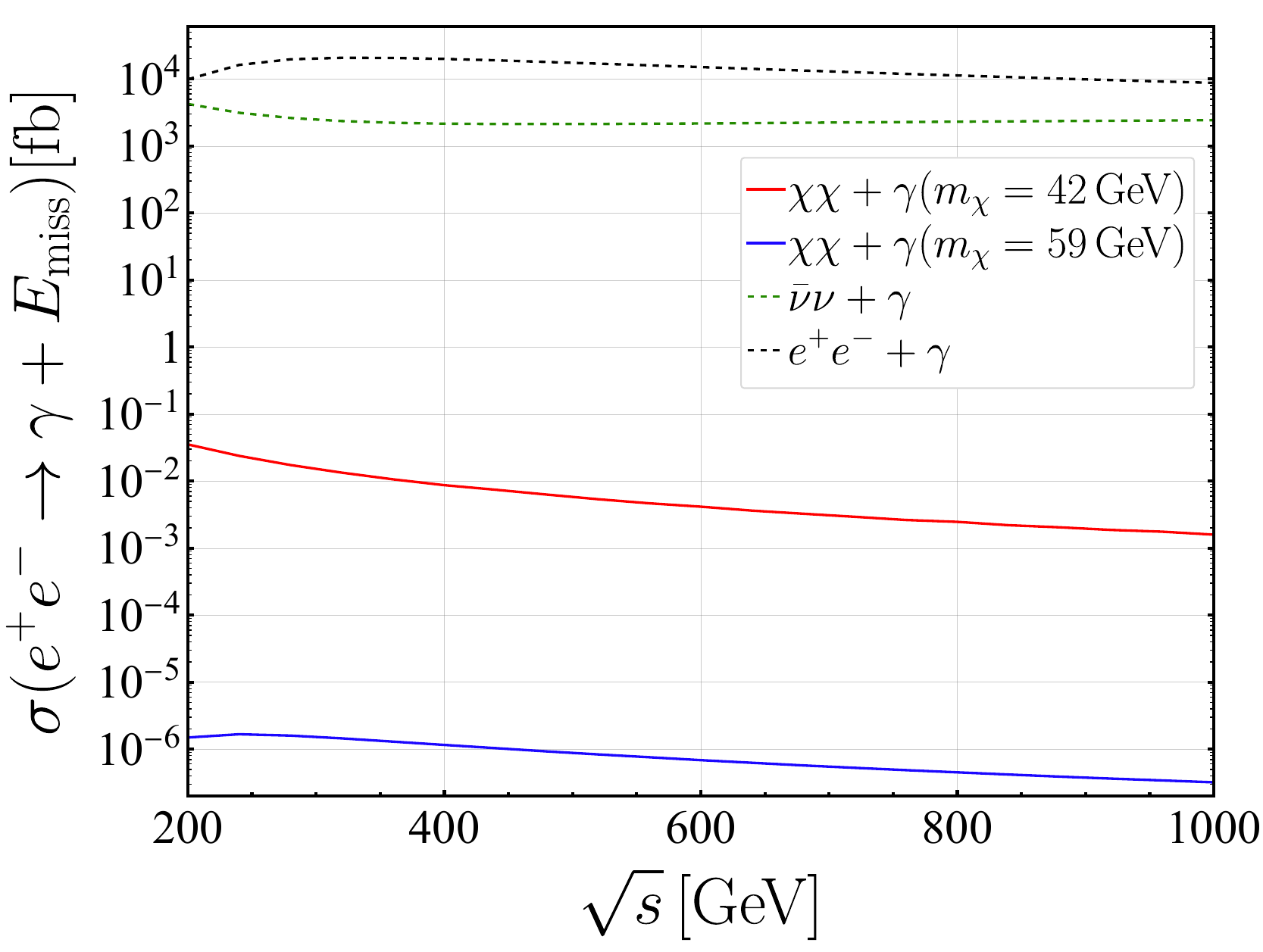}
\includegraphics[width=0.48\textwidth]{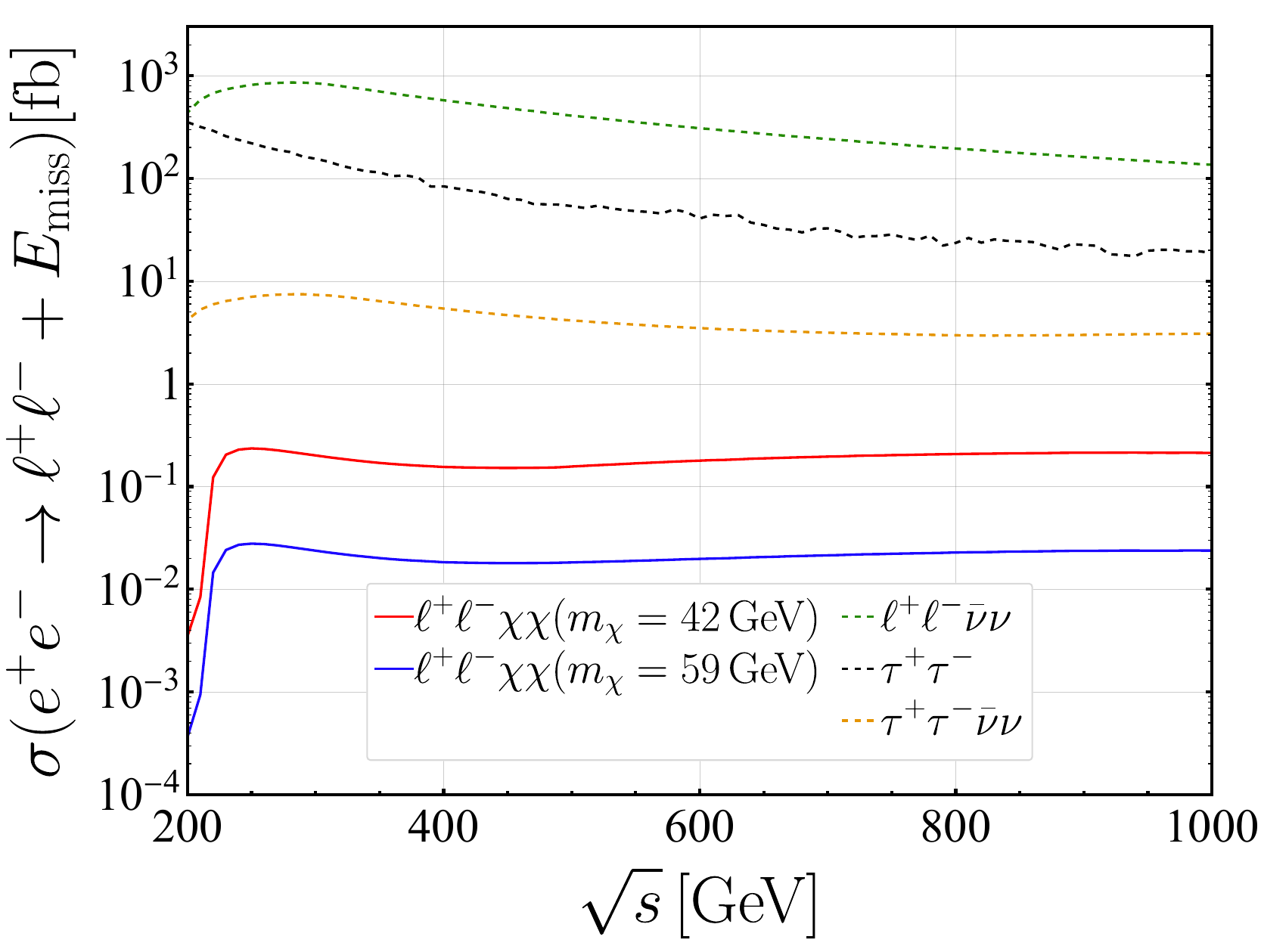}
\caption{
The left (right) panel shows the DM pair production cross sections as a function of the center-of-mass energy from the mono-photon (di-lepton) channel for two benchmark DM masses: $m_\chi = 42\,\mathrm{GeV}$ (solid red curve) and $m_\chi = 59\,\mathrm{GeV}$ (solid blue curve). The dotted curves denote the SM background channels. }
\label{fig:scanS}
\end{figure}

A comparison of the production cross section as a function of the center-of-mass energy $\sqrt{s}$ between the two channels is presented in \cref{fig:scanS}, where the left and right panels correspond to the mono-photon ($\chi\chi+\gamma$) and di-lepton ($\ell^+\ell^-\chi\chi$) channels, respectively. 
Two benchmark DM masses $m_{\chi} = 42\,\mathrm{GeV}$ and $m_{\chi} = 59\,\mathrm{GeV}$ are considered, indicated by solid red and blue curves in the figure, respectively. 
As observed from the left panel of \cref{fig:scanS}, the mono-photon cross sections are at least five orders of magnitude lower than the associated backgrounds.
This is because the mono-photon process is suppressed by the $s$-channel propagator and an additional electromagnetic coupling $\alpha_{\mathrm{em}}$, and 
meanwhile the backgrounds are also notoriously difficult to suppress~\cite{Konar:2009ae,Dreiner:2012xm,Yu:2014ula,Barman:2021hhg}.       
On the other hand, the dominant contribution to the di-lepton channel originates from the vector boson fusion (VBF) process, which produces a Higgs boson followed by its subsequent decay into $\chi\chi$ pair (see the last diagram in~\cref{fig:dileptonsignal}).
The signal cross sections rise when $\sqrt{s}$ approaches $2\,M_h$. Notably, they remain sizable at higher energies; at the same time, the background is slightly reduced from the low-$\sqrt{s}$ regime. 
Therefore, we focus on the di-lepton channel in the following discussion.

Exploiting the enhanced di-lepton signal production, we simulate the di-lepton plus missing energy signature and the associated background processes to assess the detectability of DM, focusing on the high center-of-mass energy scenario ($\sqrt{s}=1$ TeV) at the ILC~\cite{Adolphsen:2013kya}.
The pre-selection of the leptons in the simulation is required to have a transverse momentum of $P^\ell_T > 10\,\mathrm{GeV}$, guaranteeing measurable energy deposition in the electromagnetic calorimeter (ECAL). Furthermore, a loose pseudo-rapidity requirement of $|\eta_\ell| < 3.0$ is imposed on the di-lepton states to match the nominal angular acceptance of the detector ($5.7^\circ\,<\theta<174.3^\circ$).
To maximize the signal-to-background ratio, we select the following four kinematic cuts based on the event distributions at three DM benchmark points $m_\chi = 48,\,59,\,61\,\mathrm{GeV}$, as shown in~\cref{fig:BPcutfinal}:
\begin{figure}[t]
\centering
\includegraphics[width=0.48\linewidth]{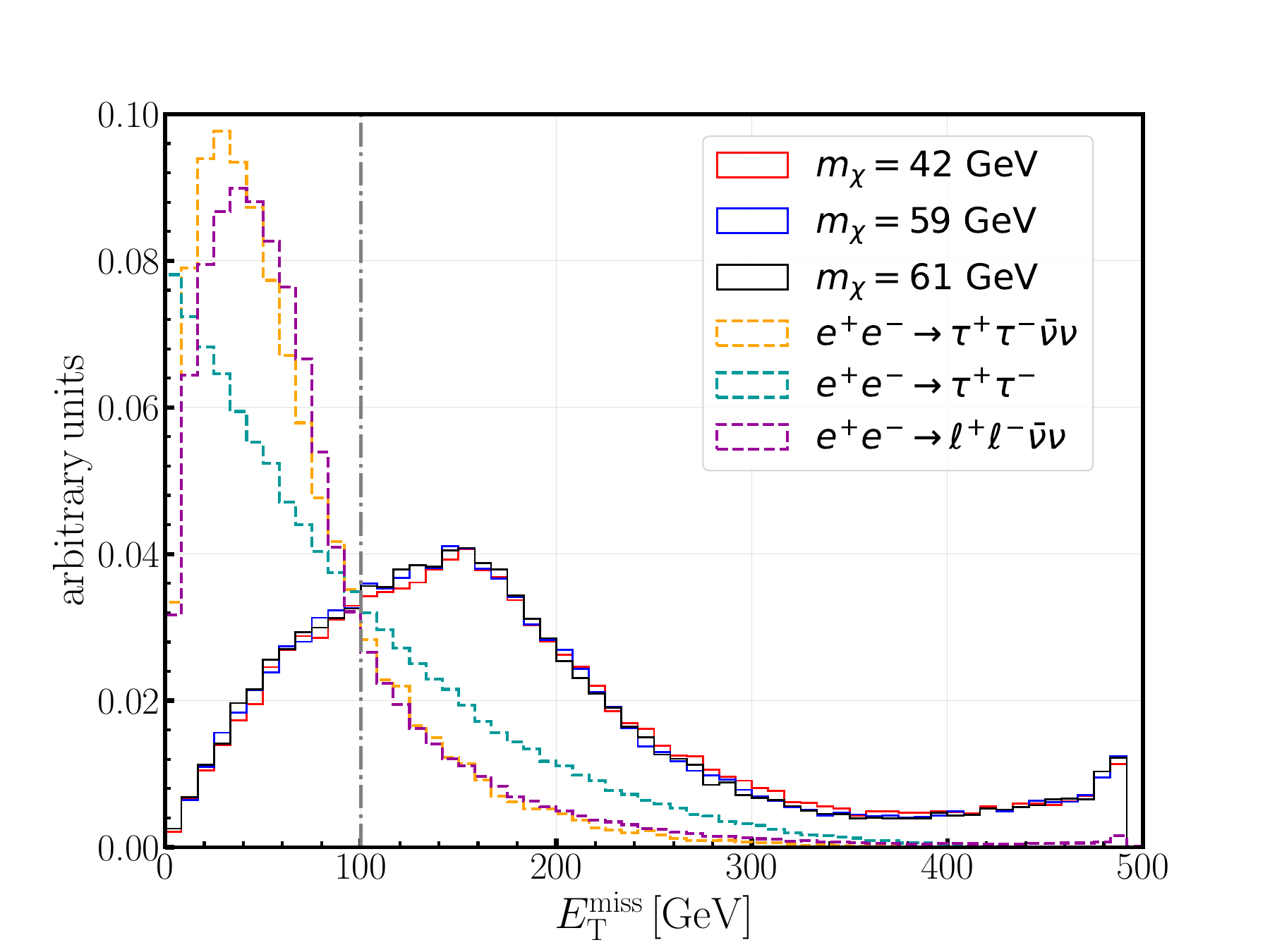}
\includegraphics[width=0.48\linewidth]{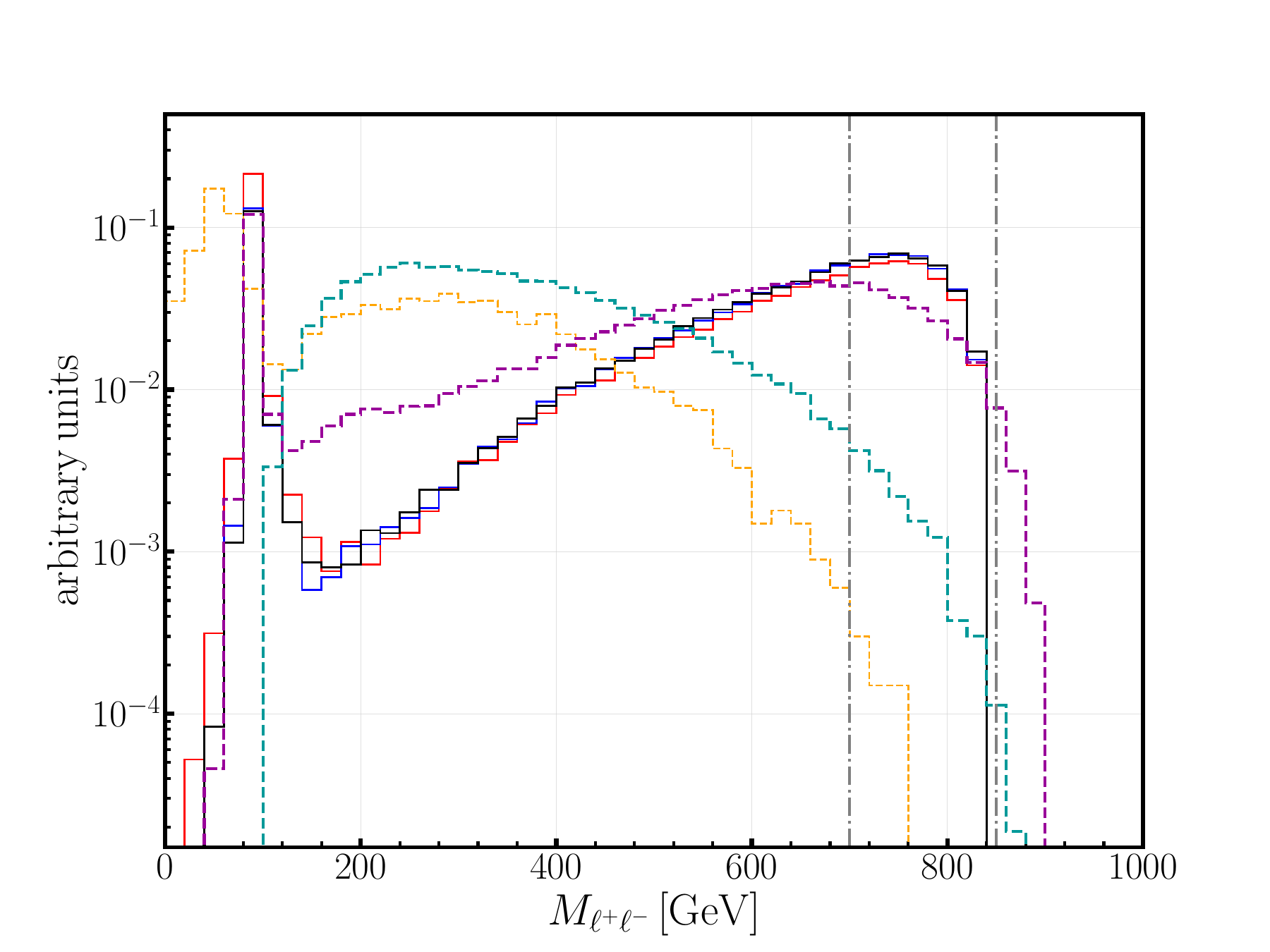}
\\
\includegraphics[width=0.48\linewidth]{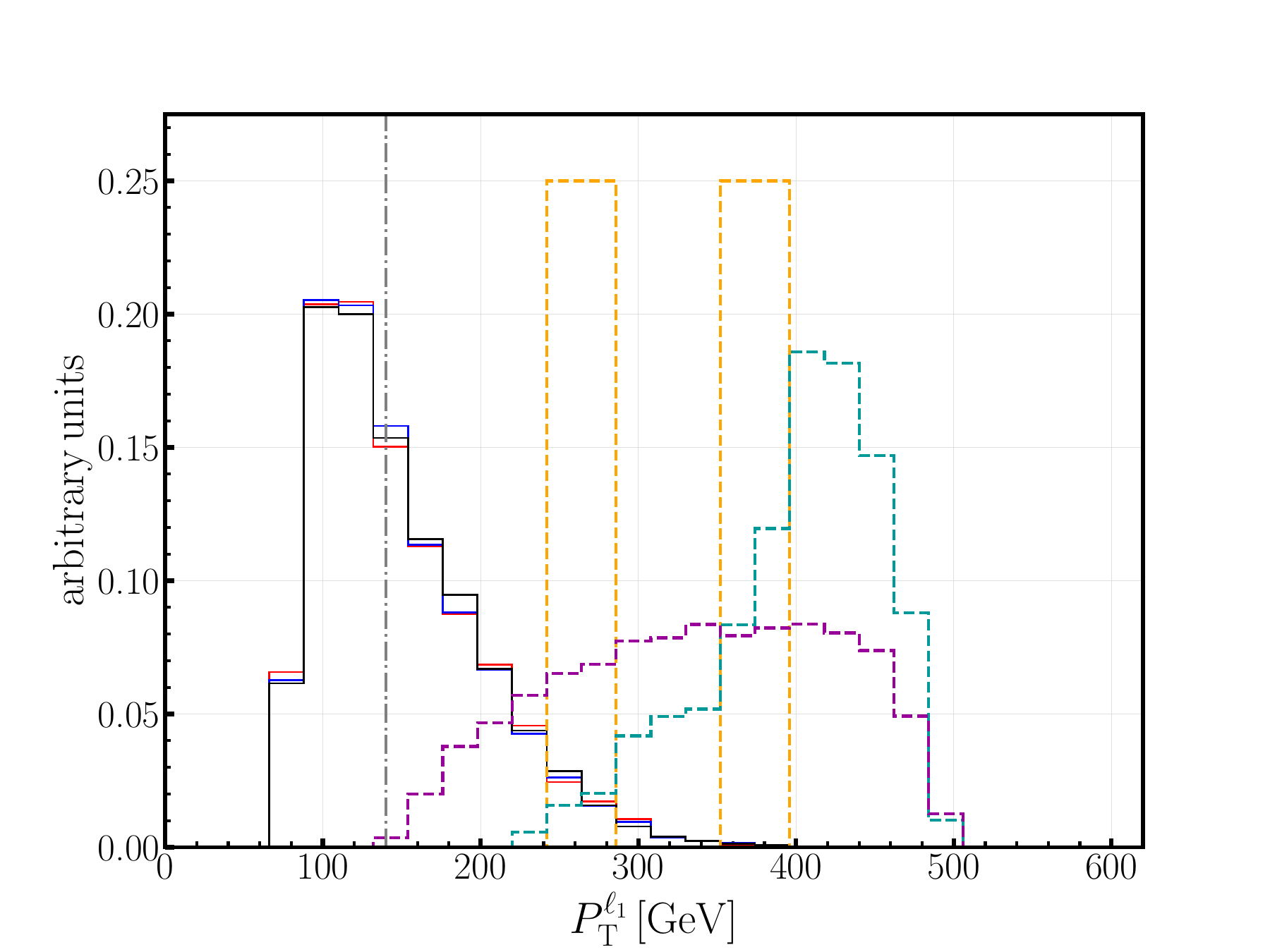} 
\includegraphics[width=0.48\linewidth]{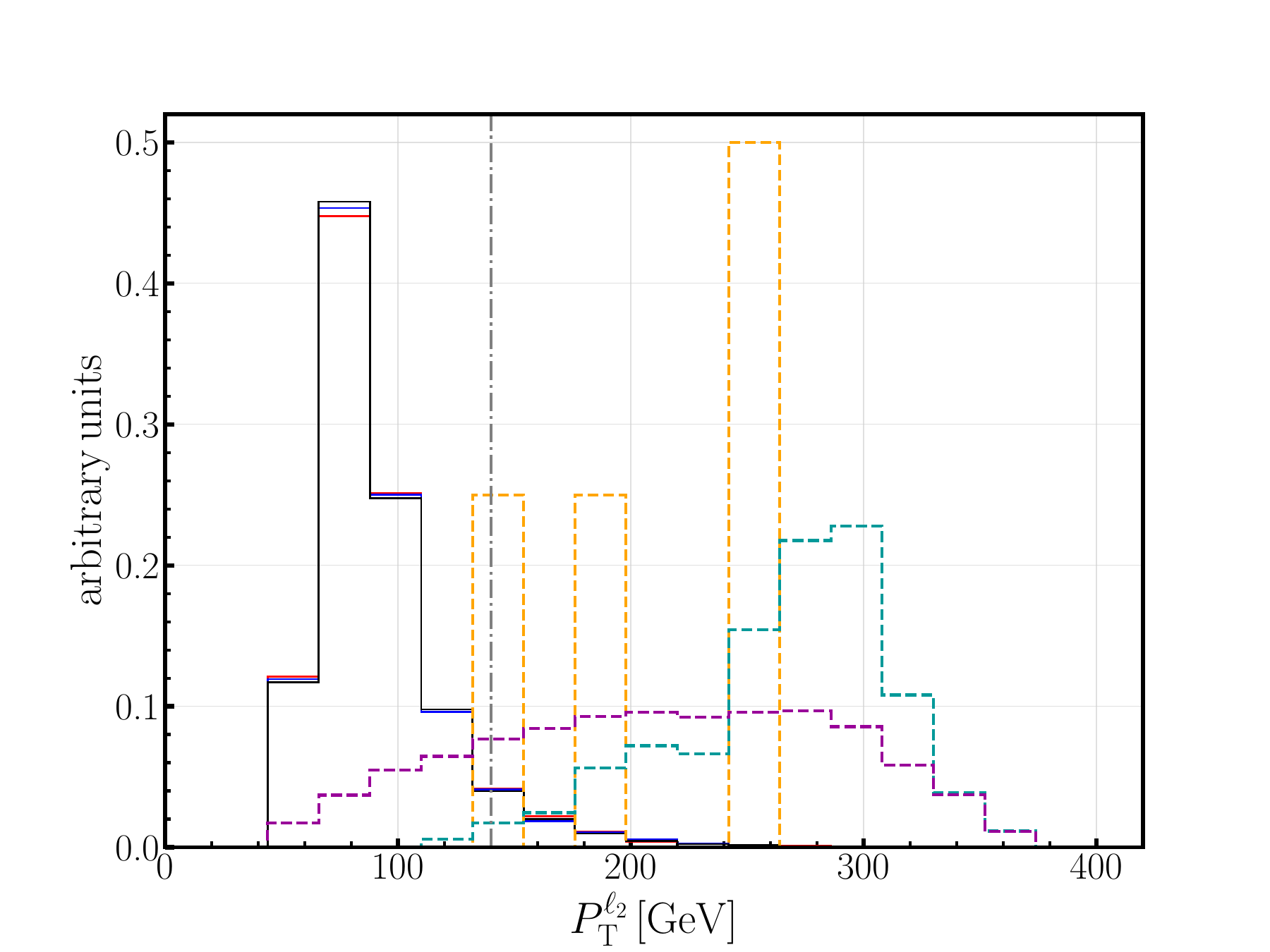}
\caption{
The signal and background event distributions in various kinematic variables. 
The top-left (-right) panel shows the event distribution of missing transverse energy $E_T^{\rm miss}$ 
(di-lepton invariant mass $M_{\ell^+\ell^-}$) before \textbf{Cut 2} (\textbf{Cut 3}). The bottom-left (-right) panel displays the leading (sub-leading) lepton transverse momentum $P_T^{\ell_{1}}$ ($P_T^{\ell_{2}}$) before \textbf{Cut 4}. Selection cuts are indicated by vertical gray dot-dashed lines.}
\label{fig:BPcutfinal}
\end{figure}
\begin{itemize}
\item \textbf{Cut\,1:} Select events with exactly two charged leptons ($N_{\ell}=2$, with $\ell=e\,\text{~or~}\mu$) whose transverse momentum $P^\ell_T>20\,\mathrm{GeV}$ and the pseudorapidity $|\eta_\ell|<2.44$ are required. The conditions for $P_T^\ell$ and $|\eta_\ell|$ are set to match the extremely high lepton identification efficiency of the ILC detector at large transverse momentum~\cite{Behnke:2013lya}\footnote{The detector has an inefficient region between the calorimeter barrel and endcap, where the tagging efficiency of the electron drops below 50\,\% for $30^\circ < \theta < 40^\circ$. However, our simulation shows that the impact from this efficiency reduction is negligible.}.
\item\textbf{Cut\,2:} Reject ${E}^{\text{miss}}_T < 100\,\mathrm{GeV}$. The top-left panel of~\cref{fig:BPcutfinal} shows the missing transverse energy distribution for signals and backgrounds after \textbf{Cut\,1}. Background events typically have a small ${E}^{\text{miss}}_T$ due to multiple low-energy neutrinos. In contrast, DM production yields a larger ${E}^{\text{miss}}_T$ from heavier invisible particles, so requiring ${E}^{\text{miss}}_T > 100\,\mathrm{GeV}$ effectively suppresses the background.
\item\textbf{Cut\,3:} Select $700 \,\mathrm{GeV}<M_{\ell^+\ell^-} <850\,\mathrm{GeV}$. The top-right panel of~\cref{fig:BPcutfinal} shows the di-lepton invariant mass $M_{\ell^+\ell^-}$ distribution after \textbf{Cut 2}. The signal concentrates in the high-mass region because the $t$-channel VBF production emits energetic leptons. Although a signal peak appears around $90$–$100\,\mathrm{GeV}$, where the SM background dominates due to $Z$-boson decays.
Therefore, this region is not suitable for signal discrimination.
\item\textbf{Cut\,4:} Select events with $P^{\ell_1}_T<140\,\mathrm{GeV}$ for the leading lepton state (i.e., one with the highest $P_T$). The bottom panels of~\cref{fig:BPcutfinal} show the leading $P_T^{\ell_1}$ (sub-leading $P_T^{\ell_2}$) distributions in the left (right) panel after \textbf{Cut 3}. Since the energetic leptons are close to the beam axis, the leading and sub-leading leptons tend to have low $P_T$ due to large pseudorapidity. This feature helps discriminate the signal from the remaining backgrounds, especially the $\ell^+\ell^- \bar{\nu}\nu$ process (shown by the purple dashed curve in the bottom-left panel), achieving significant background suppression while preserving the signal.
\end{itemize}

\begin{table}[t]
    \centering
    \resizebox{\textwidth}{!}{
    \begin{tabular}{c|c|c|c|c|c|c}
    \hline
    \hline
          & $\ell^+\ell^-\bar{\nu}\nu$  & $\tau^+\tau^-$ & $\tau^+\tau^- \bar{\nu}\nu$ & BP1 ($m_\chi = 42 \,\mathrm{GeV}$) & BP2 ($m_\chi = 59 \,\mathrm{GeV}$) & BP3 ($m_\chi = 61 \,\mathrm{GeV}$) \\
          &$\sigma$ & $\sigma$ &$\sigma$ & $\sigma (\mathcal{S})$ &$\sigma(\mathcal{S})$ & $\sigma(\mathcal{S})$ \\
          \hline 
         Pre-selection & 118.7  & 18.9 & 3.1  & 2.1$\times 10^{-1} $ (1.13) & 2.4$\times 10^{-2} $ (0.13) & 2.1$\times 10^{-3} $ (0.01)\\
         \hline 
        \textbf{Cut 1} & 50.0  & 14.5 & 0.56  & 1.1$\times 10^{-1} $ (0.84) & 1.2$\times 10^{-2} $ (0.09) & 1.0$\times 10^{-3} $ (0.01)\\
         \hline 
       \textbf{Cut 2} & 10.3 & 5.0 & 0.1  & 8.2$\times 10^{-2} $ (1.31) & 8.6$\times 10^{-3} $ (0.14) & {7.3$\times 10^{-4} $ (0.01)}\\
       \hline 
 \textbf{Cut 3} & {2.3} & {6.6$\times 10^{-2}$}& {6.3$\times 10^{-5}$}  & {2.7$\times 10^{-2} $ (1.12) } & {3.2$\times10^{-3}$ (0.13)} & {2.7$\times 10^{-4} $ (0.01) }\\
 \hline 
\textbf{Cut 4} &{2.4$\times 10^{-4} $} & {0.0}  & {0.0} & {1.5$\times 10^{-2} $ (7.60) } & {1.7$\times 10^{-3}$ (2.48) }  & {1.4$\times 10^{-4} $ (0.47) } \\
\hline 
\end{tabular} }
\caption{Cutflow for cross sections in units of fb of the background processes ($\ell^+\ell^-\bar{\nu}\nu$, $\tau^+\tau^-$, $\tau^+\tau^- \bar{\nu}\nu$) and the signal BPs at the  ILC. The number in parentheses represents the statistical significance $\mathcal{S}$ after each cut.}
\label{tab:BPandbackground}
\end{table}

\begin{figure}[t]
\centering
\includegraphics[width=0.5\linewidth]{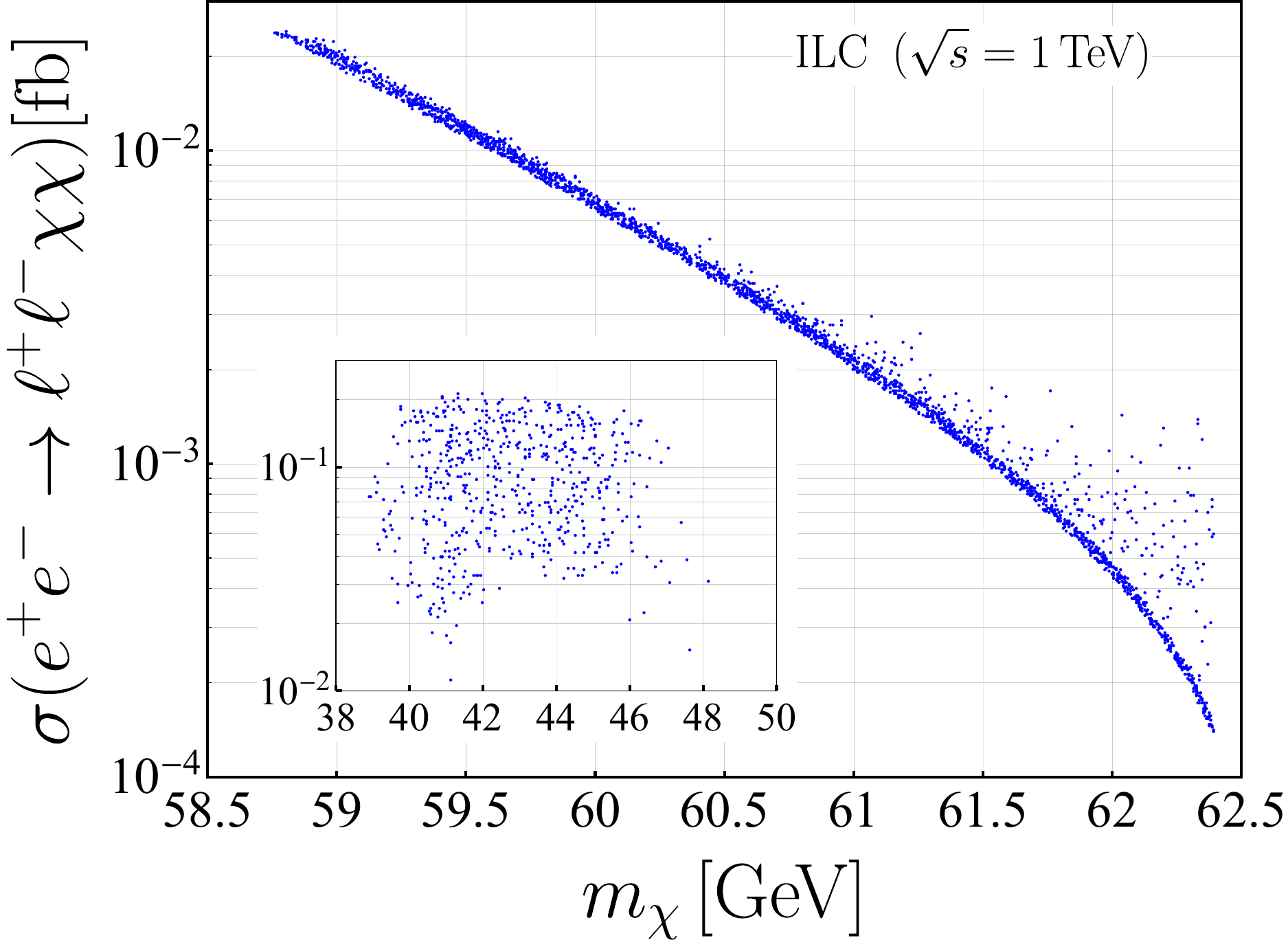}
\includegraphics[width=0.478\linewidth]{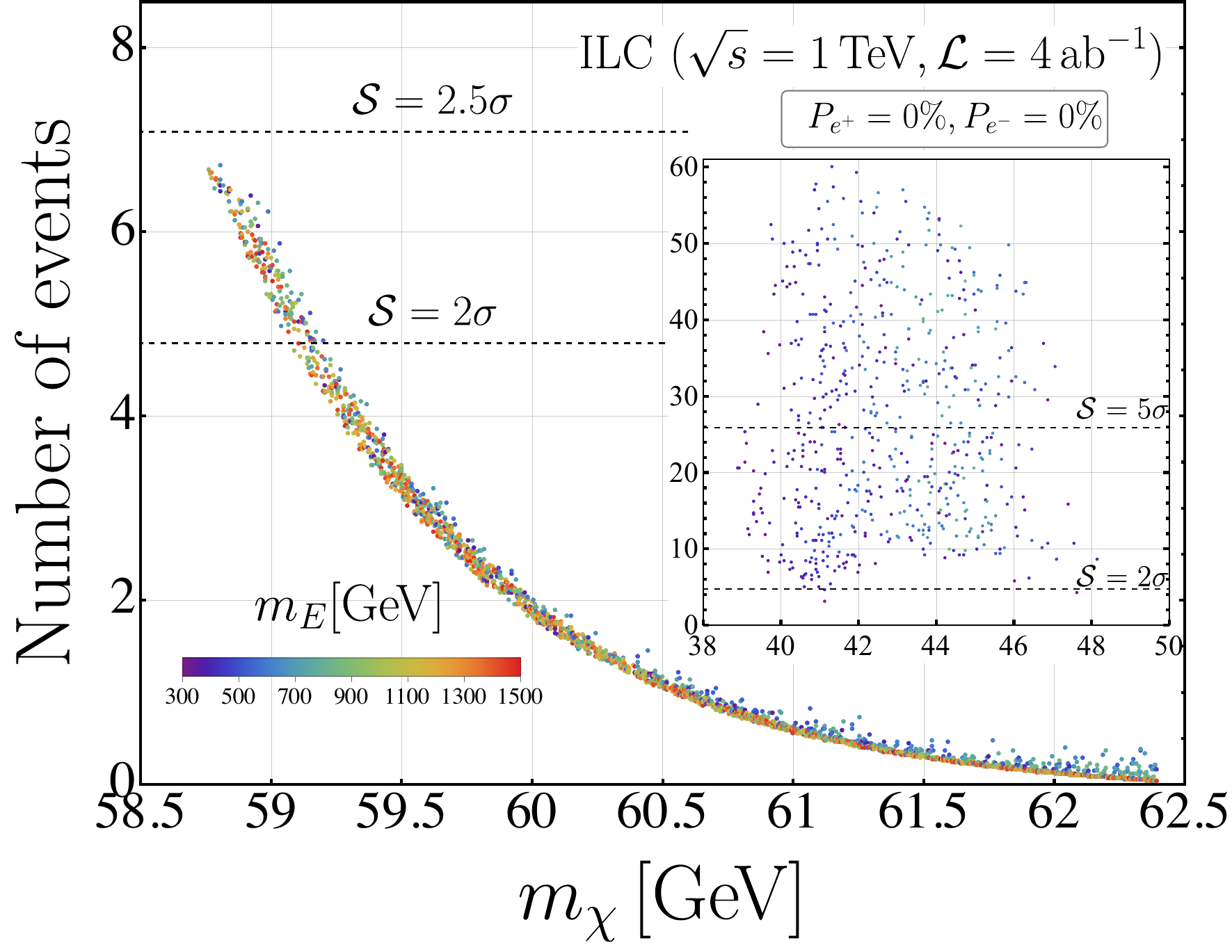}
\caption{
The left panel shows the production cross section before the selection cuts at the ILC, while the right panel displays the number of signal events after applying all four selection cuts. 
The horizontal black dashed lines show the $2\,\sigma$, $2.5\,\sigma$, and $5\,\sigma$ significance levels from Eq.~\eqref{eq:sig}.}
\label{fig:DMvsEvents}
\end{figure}

Following the luminosity benchmark for $500\,\mathrm{GeV}$ collisions at the ILC in Refs.~\cite{Zarnecki:2020ics,Fujii:2017vwa},  we extrapolate the same integrated luminosity $\mathcal{L}=4\,\text{ab}^{-1}$ to the 1 TeV scenario. 
The corresponding cross-section cutflow for these benchmark points (BPs) is shown in~\cref{tab:BPandbackground}. 
The statistical significance $\mathcal{S}$ is calculated as follows:
\begin{align}
\label{eq:sig}
\mathcal{S} = \frac{N_S}{\sqrt{N_S+N_B}},
\end{align}
where $N_S(N_B)$ denotes the number of signal (background) events that survive each selection cut.
In the table, after applying \textbf{Cut 4}, 
the signal selection efficiency reaches about $6\,\%\sim7\,\%$ while the backgrounds are significantly suppressed. 
The relatively low efficiency arises from the stringent invariant mass requirement $700\,\mathrm{GeV}<M_{\ell^+\ell^-}<850\,\mathrm{GeV}$, shown in the top-right panel of \cref{fig:BPcutfinal}. 
In the left panel of~\cref{fig:DMvsEvents}, 
we show the production cross section (in fb) versus the DM mass $m_\chi$ before all the above selection cuts. For comparison, we also show the excluded DM region around $M_Z/2$ in the insets.
The resulting signal events and statistical significance $\mathcal{S}$ after all cuts are shown in the right panel, with a conservative selection efficiency of 7\% (6.5\,\% for the inset DM region). The color bar represents the value of the vector-like fermion mass $m_E$.
As observed, the cross section or number of events decreases rapidly as the DM mass increases. 
The insets reveal that the excluded points from current experiments on the SD DM-neutron cross section yield higher event counts, with statistical significances exceeding the $5\,\sigma$ threshold. 
Within the surviving parameter region consistent with current direct detection experiments, only the mass range, $58.7\,\mathrm{GeV} \lesssim m_\chi \lesssim 59.3 \,\mathrm{GeV}$, has a discovery potential, which can achieve statistical significances above $2\,\sigma$.

\begin{table}[t]
\centering
\resizebox{0.6\textwidth}{!}{
\begin{tabular}{c|c|c|c|c}
\hline
$(P_{e^+},P_{e^-})$ & $\ell^+\ell^-\bar{\nu}\nu$  & $\tau^+\tau^-$ & $\tau^+\tau^- \bar{\nu}\nu$ & BP3 ($m_\chi = 59 \,\mathrm{GeV}$)  \\
(\%,\%) &$\sigma$ & $\sigma$ &$\sigma$ & $\sigma $  \\
\hline 
(-20,-60) &  150.3 & 17.9 & 4.0 & 2.6$\times 10^{-2} $ \\
\hline 
(-20,60) & 40.5 & 20.2 & 1.1 & 2.0$\times 10^{-2} $ \\
\hline
(0,0) & 118.7  & 18.9 & 3.1 & 2.4$\times 10^{-2} $ \\
\hline 
(20,-60) & 224.6  & 22.0 & 5.9  & 2.8$\times 10^{-2} $ \\
\hline 
(20,60) & 58.0  & 16.1 & 1.6 & 2.2$\times 10^{-2} $ \\
\hline
\end{tabular}
}
\caption{Cross sections of backgrounds ($\ell^+\ell^-\bar{\nu}\nu$, $\tau^+\tau^-$ , $\tau^+\tau^- \bar{\nu}\nu$) and the signal BP3 (in fb) under various polarization configurations before 
{\bf Cuts 1,\,2,\,3,\,4}.  }
\label{tab:polarvalues}
\end{table}

\begin{figure}[t]
\centering
\includegraphics[width=0.504\linewidth]{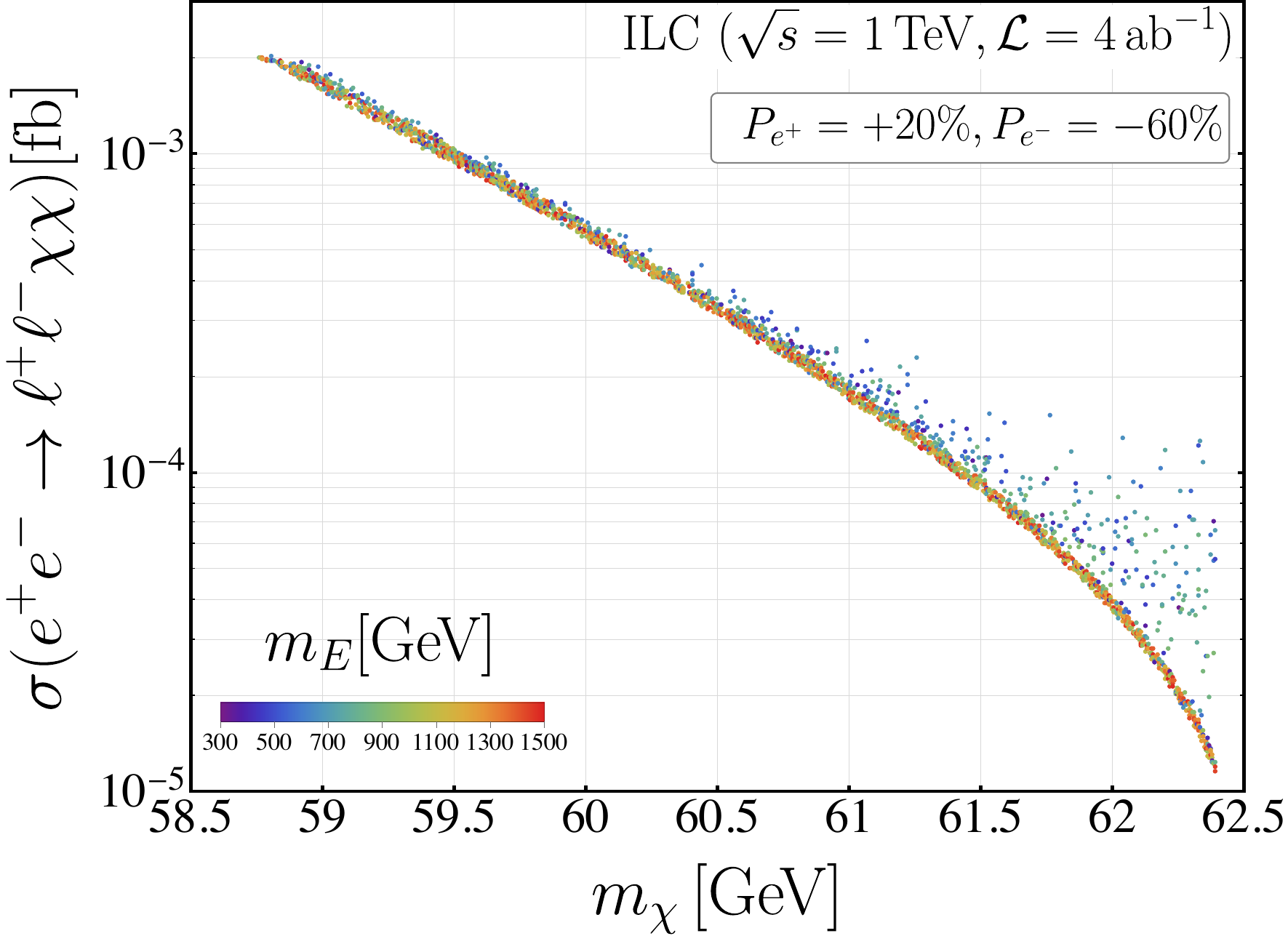}
\includegraphics[width=0.478\linewidth]{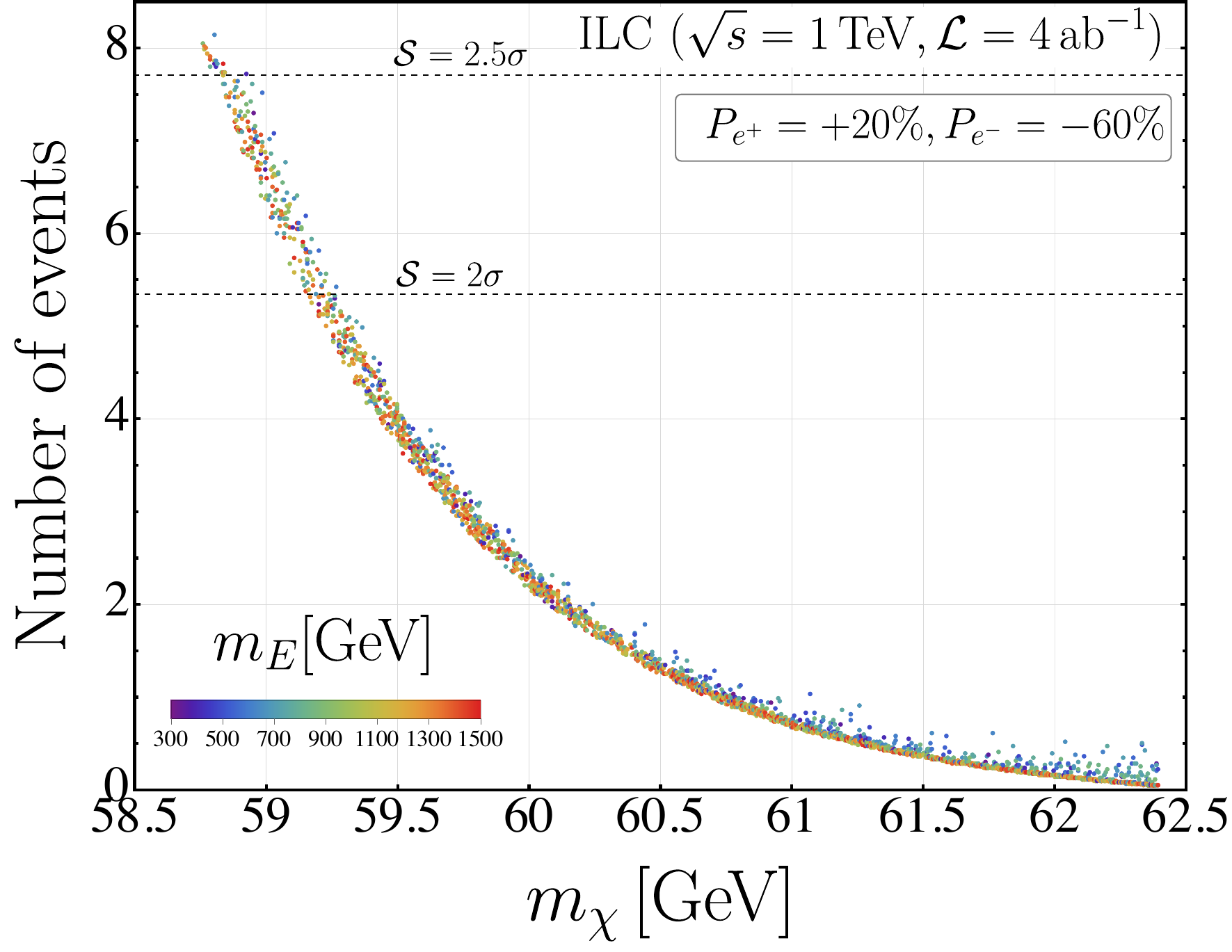}
\caption{Left: The production cross section after applying {\bf Cut 1,2,3,4}. 
Right: Final number of events as a function of DM mass.}
\label{fig:cutfinalpolar}
\end{figure}

Finally, we consider improving the statistical significance by employing polarized lepton beams. For the case of ILC energy $\sqrt{s}=500\,\rm GeV$, the positron and electron beams are expected to reach polarizations of about 30\,\% and 80\,\%, respectively~\cite{Bambade:2019fyw}. 
Adopting a conservative configuration of 20\,\% (60\,\%) polarized positron (electron) at $\sqrt{s}=1$ TeV, we compare four polarization configurations with the unpolarized case in~\cref{tab:polarvalues}. 
As seen in the table, the 
$(P_{e^+},\,P_{e^-})=(+20\,\%,\,+60\,\%)$ polarization case strongly suppresses the dominant $e^+ e^- \to \ell\ell \bar{\nu}\nu$ background, as this process is mediated by $W^+W^-$ bosons while the initial right-handed electrons do not couple to $W$ bosons, resulting in a reduced background.
However, this background suppression effect is marginal compared to selection cuts in \cref{tab:BPandbackground}.  
Therefore, we adopt the beam polarization $(P_{e^+},\,P_{e^-})=(+20\,\%,\,-60\,\%)$ in the analysis, as it yields a larger signal cross section. 
The final result with this polarization choice after all selection cuts is given in \cref{fig:cutfinalpolar}.
The left panel shows the production cross section vs the DM mass, while the right panel displays the number of events, with black dashed lines indicating $2\,\sigma$ and $2.5\,\sigma$ statistical significances.
This selection strategy achieves $\sim 2.5\,\sigma$ significance for $m_\chi \sim 59\,\mathrm{GeV}$, which is slightly improved over the unpolarized case shown in \cref{fig:DMvsEvents}.
In summary, the collider search complements the DM direct detection in \cref{sec:numerical-scan}, enabling the lower DM mass region $58.7\,\mathrm{GeV} \lesssim m_\chi \lesssim 59.3 \,\mathrm{GeV}$ to be probed jointly.

\section{Conclusion}
\label{sec:conclu}

In this paper, we systematically explored the phenomenology of fermionic DM candidate in the scotogenic inverse seesaw model. 
By synthesizing constraints from neutrino oscillation data, LFV processes, DM relic density, and current DM direct detection, we uncovered a viable DM window in the mass range of $58\,{\rm GeV} \lesssim m_\chi \lesssim 63\,{\rm GeV}$. 
We demonstrated that the window can be further tested by next-generation ton-scale DM experiments and future ILC. 

Our numerical scan shows that the DM relic density sets the most stringent constraints, while the bound from the LFV $\mu\to e\gamma$ is relatively weak. 
In the DM sector, the Majorana nature suppresses vector-current interactions, rendering the Higgs-mediated SI cross section subdominant to the $Z$-mediated SD cross section.
A key result of our analysis is that current experimental limits on SD DM-neutron scattering, particularly from the LZ experiment, entirely exclude the $Z$-boson resonance region. As a result, the surviving parameter space is confined to the Higgs resonance funnel ($m_\chi \sim M_h/2$). Furthermore, this remaining parameter space can be fully probed by future direct detection experiments such as PandaX-xT.
Interestingly, in this region the Higgs invisible decay branching ratio is enhanced to above the $\calO(10^{-3})$ level. 

At future ILC with $\sqrt{s}=1$ TeV and $\mathcal{L}=4$ ab$^{-1}$, the di-lepton plus missing energy channel ($e^+ e^- \to \ell^+\ell^-\chi\chi$) provides a more efficient probe of dark matter than the mono-photon channel ($e^+e^- \to \chi\chi+\gamma$), due to vector boson fusion enhancement of signals and weaker backgrounds. Our analysis shows that the dominant backgrounds in the di-lepton plus missing energy channel can be effectively suppressed through kinematic variable cuts on missing transverse energy ($E^{\mathrm{miss}}_{T}$), di-lepton invariant mass ($M_{\ell^+\ell^-}$), and the leading lepton transverse momentum ($P^{\ell_1}_{T}$). With these cut selections, the DM mass range $58.7\,\mathrm{GeV} \lesssim m_\chi \lesssim 59.3\,\mathrm{GeV}$ can be tested at a statistical significance of $2\,\sigma$. The sensitivity can be further improved by employing polarized beams, increasing the significance to approximately $2.5\,\sigma$. Consequently, future lepton colliders would provide complementary probes of dark matter together with forthcoming direct detection experiments.

\acknowledgments
This work was partially supported by Grants No.\,NSFC-12305110, No.\,NSFC-12247151, No.\,NSFC-12035008, and by SCNU Young Teachers Scientific Research Foundation (No.599/672203).

\bibliography{references}{}
\bibliographystyle{JHEP}

\end{document}